\documentclass[
    prd,preprintnumbers, twocolumn, superscriptaddress, amsmath, amssymb,
    aps, floatfix, longbibliography,nofootinbib
]{revtex4-2}
\usepackage[utf8]{inputenc}

\usepackage{graphicx,xcolor,setspace}
\usepackage[
    colorlinks=true,
    linkcolor=blue,
    urlcolor=blue,
    citecolor=blue
]{hyperref}

\usepackage{bm,relsize}
\usepackage[titletoc,toc,title]{appendix}
\usepackage{titletoc}
\usepackage{titlesec}
\usepackage{slashed}
\usepackage{simpler-wick}
\usepackage{tikz-feynman}
\usepackage{placeins}

\newcommand{\beq}{\begin{equation}}
\newcommand{\eeq}{\end{equation}}
\newcommand{\bmtx}{\begin{pmatrix}}
\newcommand{\emtx}{\end{pmatrix}}

\def\xFO			{x_{{\rm FO}}}

\def\gSM		{g_{{\rm SM}}}
\def\gDM		{g_{{\rm DM}}}

\newcommand{\G}{\mathcal{G}}
\newcommand{\mdg}{m_{\rm GB}}
\newcommand{\LL}{\rm L}

\begin{document}
\preprint{LAPTH-055/23}
\title{\Large Heavy Baryon Dark Matter from $SU(N)$ Confinement: \\
\large Bubble Wall Velocity and Boundary Effects }

\author{Yann Gouttenoire}
\email{yann.gouttenoire@gmail.com}
\affiliation{School of Physics and Astronomy, Tel-Aviv University, Tel-Aviv 69978, Israel}
\author{Eric Kuflik}\email{eric.kuflik@mail.huji.ac.il}
\affiliation{Racah Institute of Physics, Hebrew University of Jerusalem, Jerusalem 91904, Israel}
\author{Di Liu}
\email{liudisy@gmail.com}
\affiliation{Laboratoire d'Annecy-le-Vieux de Physique Th\'eorique, CNRS -- USMB, BP 110 Annecy-le-Vieux, F-74941 Annecy, France}

\begin{abstract}
Confinement in $SU(N_{\rm DC})$ Yang-Mills theories is known to proceed through first-order phase transition. The wall velocity is bounded by $v_w \lesssim 10^{-6}$ due to the needed time for the substantial latent heat released during the phase transition to dissipate through Hubble expansion.
Quarks much heavier than the confinement scale can be introduced without changing the confinement dynamics. After they freeze-out, heavy quarks are squeezed into pockets of the deconfined phase until they completely annihilate with anti-quarks. We calculate the dark baryon abundance surviving annihilation, due to bound-state formation occurring both in the bulk and - for the first time - at the boundary. We find that dark baryons can be dark matter with a mass up to $10^3~\rm TeV$. We study indirect and direct detection, CMB and BBN probes, assuming portals to Higgs and neutrinos.
\end{abstract}
\maketitle

\raggedbottom

\section{INTRODUCTION}
\label{sec:intro}

The nature of Dark Matter (DM) composing $85\%$ of the matter content of the universe remains unknown. The possibility that DM is a composite state from a hidden strong sectors has been studied extensively \cite{Kribs:2016cew,Cline:2021itd}, either as glueballs \cite{Carlson:1992fn,Faraggi:2000pv,Feng:2011ik,Boddy:2014yra,Pappadopulo:2016pkp,Juknevich:2009ji,Juknevich:2009gg,Forestell:2016qhc,Forestell:2017wov,Soni:2016gzf,Soni:2016yes,Acharya:2017szw,Carenza:2022pjd,Carenza:2023shd}, pseudo Nambu-Goldstone bosons \cite{Bellazzini:2014yua,Cacciapaglia:2020kgq,Hietanen:2013fya,Hochberg:2014dra,Hochberg:2015vrg,Contino:2020god}, dark baryons \cite{Nussinov:1985xr,Strassler:2006im,Kribs:2009fy,Bai:2013xga,Pasechnik:2014ida,LSD:2014obp,Huo:2015nwa,Antipin:2014qva,Antipin:2015xia,Mitridate:2017izz,Mitridate:2017oky,Berryman:2017twh,Dondi:2019olm,Redi:2018muu,Garani:2021zrr}, or quark nuggets \cite{Witten:1984rs,Frieman:1990nh,Zhitnitsky:2002qa,Oaknin:2003uv,Atreya:2014sca,Gresham:2017cvl,Bai:2018vik,Bai:2018dxf,Ge:2019voa,Gross:2021qgx}.
The benefit of dark baryons lies in their intrinsic stability, which is ensured by the accidental conservation of baryon number, see e.g. \cite{Antipin:2015xia}. If quarks masses are comparable or lighter than the confinement scale $m_\mathcal{Q} \lesssim \Lambda$, then the dark baryon abundance is set by the standard freeze-out mechanism \cite{Kolb:1990vq} which leads to the prediction of a mass around $100~\rm TeV$. Scenarios for which quarks are much heavier $m_{\mathcal{Q}}\gtrsim 10^2 \Lambda$ can have richer dynamics \cite{Harigaya:2016nlg,DeLuca:2018mzn,Gross:2018zha,Geller:2018biy,Dondi:2019olm,Mitridate:2017oky,Asadi:2021pwo,Asadi:2021yml,Asadi:2022vkc} due to quarks freezing-out before confinement. Lattice simulations have shown that confinement Yang-Mills $SU(N_{\rm DC})$ gauge theory occurs through a first-order phase transition (PT) \cite{Lucini:2003zr,Lucini:2005vg,Panero:2009tv,Saito:2011fs}.
As demonstrated in \cite{Asadi:2021pwo,Asadi:2021yml}, this results in a `thermal squeezeout' scenario during which heavy quark relics are compressed into shrinking pockets of the deconfined phase until the majority re-annihilate with anti-quarks. During this process, quarks form a small fraction of baryons in the bulk of the pocket which subsequently escape through the wall boundary \cite{Asadi:2021pwo,Asadi:2021yml}, and explain DM with a mass beyond the unitarity bound $m_{\rm DM}\sim 10^2~\rm TeV$. In the present work, we introduce a new formalism to investigate the formation of bound-states directly at the wall boundary. Those boundary effects were omitted in \cite{Asadi:2021pwo,Asadi:2021yml}. They can be understood as the strongly-coupled analog to the filtered DM mechanism introduced in \cite{Baker:2019ndr,Hong:2020est,Lu:2022paj}. For composite sectors, particle production at boundaries has only been studied in the relativistic bubble wall velocity limit \cite{Baldes:2020kam,Baldes:2021aph}. In addition to introduce novel boundary effects, this analysis extends the original study in \cite{Asadi:2021pwo,Asadi:2021yml} to much smaller confining scales $\Lambda \ll \rm TeV$. At first, we study in details the bubble growth dynamics and the wall velocity, which we find to be small $v_w\sim 10^{-6}$. Next, we study in details the baryon abundance, accounting for thermal freeze-out before the PT, solving a set of coupled Boltzmann equations during the squeezing phase of the PT and finally accounting for eventual entropy injection following glueball decay after the PT. We propose analytical formulae for the dark baryon abundance which are successfully tested against numerical results. We find that boundary effects can reduce the predicted dark baryon DM mass by about one order of magnitude with respect to the prediction from \cite{Asadi:2021pwo,Asadi:2021yml}, leading to $m_{\rm DM} \sim 10^3~\rm TeV$. We introduce two possible portals with the Standard Model (SM), through the Higgs and neutrino sectors, and we study the relevant phenomenology.

\begin{figure*}[ht!]
    \centering
    \includegraphics[width=0.7\textwidth]{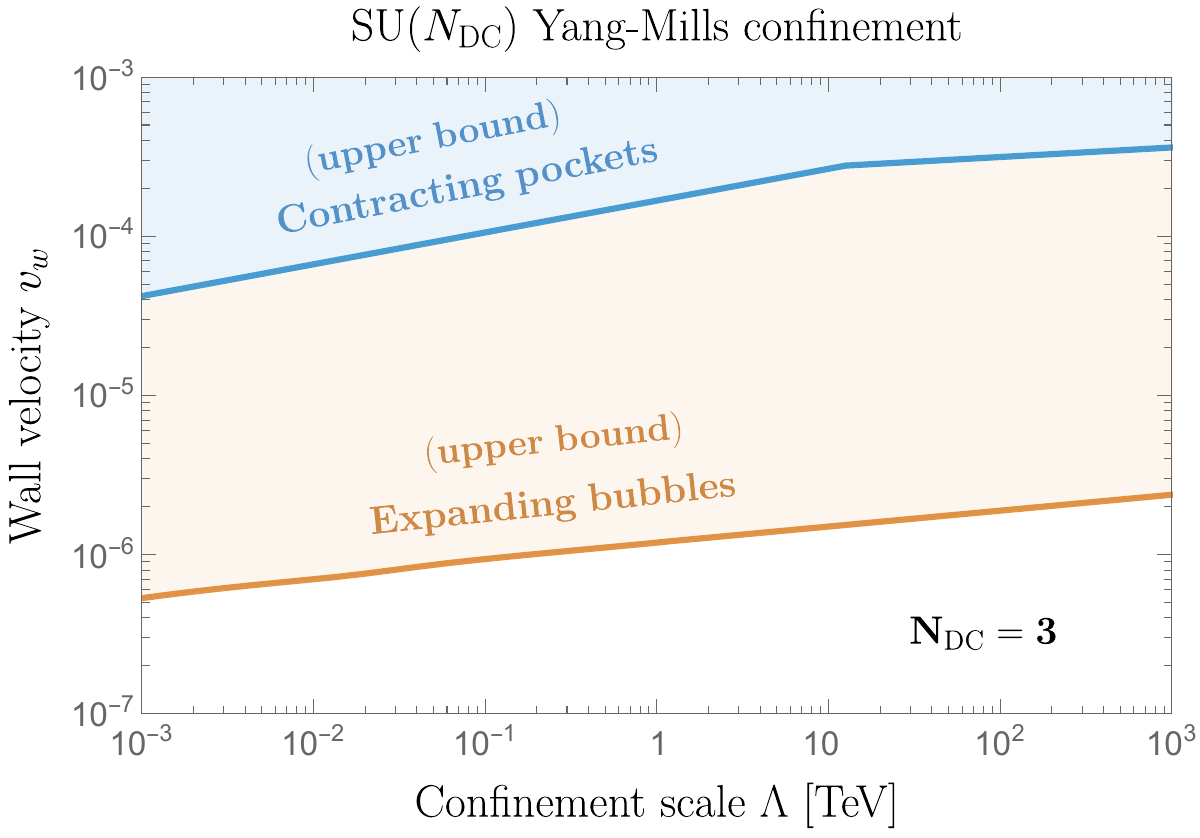}
\caption{$SU(N_{\rm DC})$ Yang-Mills phase transitions release a large latent heat $L\sim T_c^4$ while the free energy difference driving the wall motion is very small $\Delta f \sim 10^{-3}T_c^4$. This implies that bubbles can only grow after that the universe has cooled under Hubble expansion. We show here the wall velocity for expanding bubbles in Eq.~\eqref{eq:v_w_Hubble} (\textbf{Orange}) and contracting pockets in Eq.~\eqref{eq:v_w_contraction} (\textbf{Blue}) set by Hubble expansion rate. Those formulae assume that the latent heat has been transferred to the whole universe. Any accumulation around bubble walls would lead to a lower velocity. Also additional pressure contributions due to particle squeezing, e.g. heavy quarks \cite{Asadi:2021pwo}, could further slow down the pocket contraction rate (Blue only). For those reasons, we consider the values predicted in Eqs.~\eqref{eq:v_w_Hubble} and \eqref{eq:v_w_contraction} as upper limits.}
    \label{fig:wall_velocity_upper_bound}
\end{figure*}

\section{BUBBLE WALL VELOCITY}
Results of this section apply to all Yang-Mills PT, independently of the existence of heavy quarks.\\
\subsection{Bubble nucleation}
$SU(N_{\rm DC})$ Yang-Mills phase transition are known to be first order for $N_{\rm DC}\geq 3$ \cite{Lucini:2003zr,Lucini:2005vg,Panero:2009tv}. In presence of quarks, the transition remains first-order if the later are much heavier than the confinement scale \cite{Brown:1990ev,Saito:2011fs,Aarts:2023vsf}, of the order $m_{\mathcal{Q}} \gtrsim 10^2\Lambda$ \cite{Saito:2011fs}.  The tunneling probability can be inferred from lattice simulations in the thin-wall limit \cite{Fuller:1987ue}. In App.~\ref{sec:thermo}, one finds that the nucleation takes place very close to the critical temperature $T_c\simeq \Lambda$, below which the transition is energetically allowed:
\begin{equation}
    \frac{T_c-T}{T_c}~\simeq ~7\times 10^{-4}.
\end{equation}
\textbf{Bubble expansion.}
The conversion of gluons with energy density $\propto \mathcal{O}(N_{\rm DC}^2)$ into glueball with energy density $\propto \mathcal{O}(N_{\rm DC}^0)$ releases a large latent heat $L=\mathcal{O}(N_{\rm DC}^2)$. For $N_{\rm DC}=3$, lattice simulations find $L\simeq 1.413T_c^4$ \cite{Lucini:2005vg}.
If the gluons and SM exchange energy fast enough, then the SM can transport heat to the other side of the wall, leading the universe to be instantaneously reheated as bubble walls expand. If bubbles expand too fast, the latent heat released would reheat the universe above $T_c$. This implies that the bubble wall velocity is limited by the rate of Hubble expansion of the universe \cite{Asadi:2021pwo}. One finds (see App.~\ref{app:bubble_growth} for derivation):
\begin{equation}
\label{eq:v_w_Hubble}
    v_w \simeq 100 HR_0\left(\frac{T_c^4}{\LL}\right)\left(\frac{g_*}{100}\right)\left(\frac{R_0}{R}\right)^2,
\end{equation}
where $R$ is the bubble radius at a given time and $R_0=10^{-6}\left({M_{\rm pl}/\Lambda}\right)^{0.9}/\Lambda$ is the bubble radius at percolation, whose value is determined from numerical simulations \cite{Asadi:2021pwo}. In fact, Eq.~\eqref{eq:v_w_Hubble} can be understood as an upper limit which is only reached if the universe temperature has enough time to homogenize. Instead, inhomogeneity in the temperature of the plasma surrounding bubbles would cause additional heating at the wall and would lead to a lower wall velocity. The bubble wall velocity decreases as bubble walls expand, until it reaches $v_w \simeq 10^{-5}(\Lambda/100\rm TeV)^{0.1}$ for $R\simeq R_0$. 
When bubbles percolate, the slow rate of bubble wall expansion can leave the time for small bubbles to coalesce into larger bubbles in order to minimize the total surface tension \cite{Witten:1984rs}. Depending whether coalescence is fast enough, the bubble radius at the beginning of the pocket contraction stage is
\begin{equation}
\label{eq:Ri}
    R_i =\textrm{Min}\left[R_0,~R_1\right],
\end{equation}
with $R_1\Lambda=0.002(M_{\rm pl}/\Lambda)^{2/3}$ \cite{Asadi:2021pwo} and $R_0$ given below Eq.~\eqref{eq:v_w_Hubble}.\\
\subsection{Pocket contraction.}
After that bubbles percolated and eventually coalesced, pockets of deconfined phase form and begin to shrink. 
In App.~\ref{app:pocket_contraction}, we show that during the pocket contraction stage, the bubble wall velocity is again controlled by the Hubble expansion rate. However, instead of Eq.~\eqref{eq:v_w_Hubble} it is constant with radius and given by:
\begin{equation}
\label{eq:v_w_contraction}
v_w \simeq 4(HR_i)^{3/5} \simeq \textrm{Min}\left[v_0,~v_1\right],
\end{equation}
where $v_0\simeq 2\times 10^{-3}\left({\Lambda/M_{\rm pl}}\right)^{0.06}$ and $v_1\simeq 0.2\left( {\Lambda}{M_{\rm pl}}\right)^{1/5}$ for $R_i = R_0$ and $R_i=R_1$ respectively. For $\Lambda \simeq \rm TeV$, we find $v_0\simeq v_1\simeq 2 \times 10^{-4}$. The wall velocity of the expanding bubbles and contracting pockets are plotted in Fig.~\ref{fig:wall_velocity_upper_bound}.
\begin{figure*}[ht!]
    \centering \includegraphics[width=0.7\textwidth]{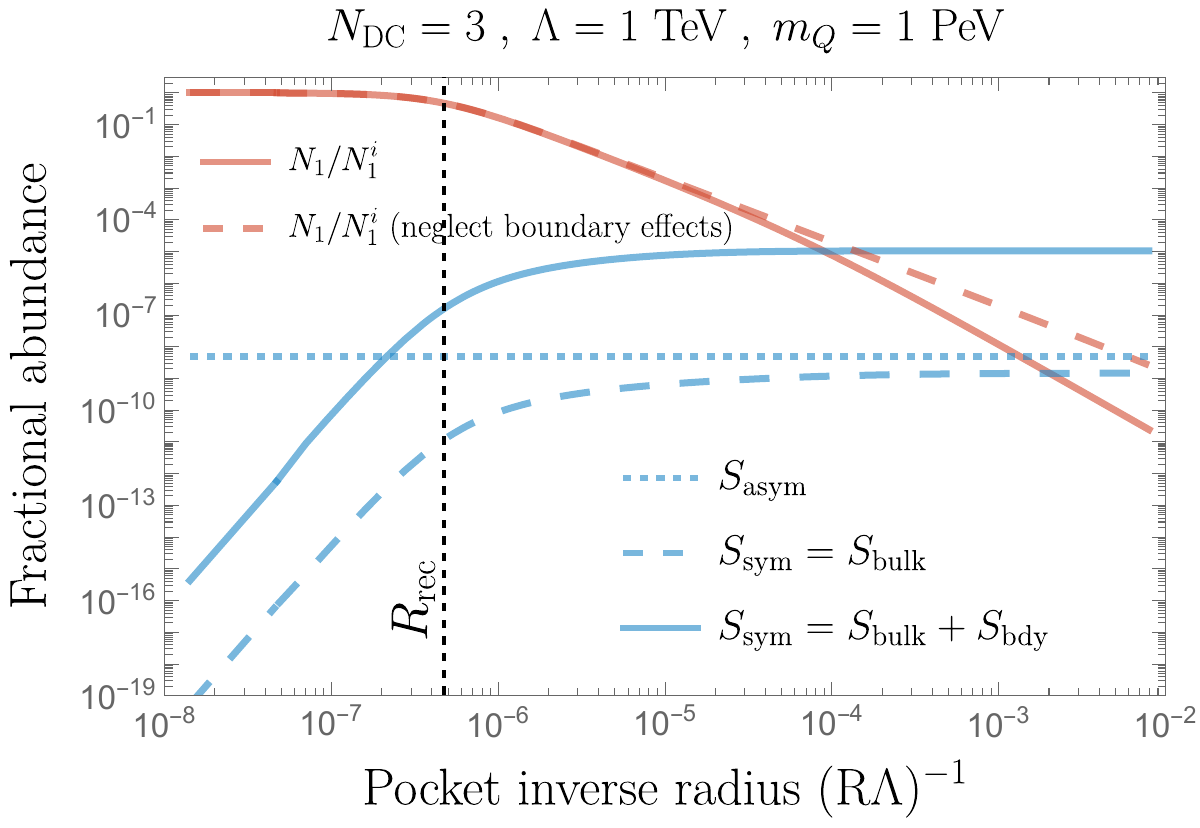}
    \caption{Evolution of the fractional number of quark inside (\textbf{Red}) and outside (\textbf{Blue}) the pocket during its contraction, obtained from numerically integrating the full system of Boltzmann equations. Below the recoupling radius $R\lesssim R_{\rm rec}$, quarks annihilate with antiquarks. Quarks surviving annihilation are the asymmetric fraction (\textbf{Blue dotted}), the baryons formed in the bulk and escaping through the boundary (\textbf{Blue dashed}) and the baryons directly formed at the boundary (\textbf{Blue solid}). The formation of baryon directly at the boundary, which is the novelty of this work, is the dominant component.}
    \label{fig:fractional_abundance}
\end{figure*}

\section{QUARK FREEZE-OUT}
\label{sec:quark_FO}
 At high temperatures, the plasma is in the deconfined phase of $SU(N_{\rm DC})$. The quarks evolve like standard thermal relics: they begin in equilibrium with the bath, become non-relativistic, and then freezeout. The abundance $Y\equiv n/s$ following freezeout of the quark annihilations can be estimated using the standard instantaneous freezeout (see App.~\ref{sec:quark_FO} for details):
 \begin{equation}
 \label{eq:Yqinf}
 Y_\mathcal{Q}^\infty \simeq \frac{H}{{s}\left<\sigma_{\rm ann} v\right>}\Bigg|_{T=T_{\rm FO}} \simeq   {0.24 \frac{x_{\rm FO}}{g_{*}^{1/2}}\frac{m_{\mathcal{Q}}}{\zeta\alpha_{\rm DC}^2 M_{\rm pl} }},
 \end{equation}
where  $s$ is the entropy density, $H$ is the Hubble rate, $g_*$ is the number of relativistic dofs, $M_{\rm pl}$ is the reduced Planck mass. We assume no quark asymmetry such that the number of anti-quarks is $ Y_{\overline{\mathcal{Q}}}^\infty= Y_\mathcal{Q}^\infty$. The freeze-out temperature is given by $x_{\rm FO} \equiv m_{\mathcal{Q}}/T_{\rm FO}\simeq \textrm{log}\left[0.19M_{\rm pl}m_{\mathcal{Q}}\left<\sigma v\right> g_{\mathcal{Q}}/g_*^{1/2}/x_{\rm FO}^{1/2}\right] $ \cite{Kolb:1990vq} with $g_{\mathcal{Q}}=2N_{\rm DC}$. The quark anti-quark annihilation cross section to dark gluons and dark photons
\begin{equation}
\left<\sigma_{\rm ann} v\right> = \zeta  {\frac{\pi \alpha_{\rm DC}^2}{m_{\mathcal{Q}}^2}},  \label{sigmav}
\end{equation}
where $\alpha_{\rm DC}$ is the dark strong coupling experienced by the fundamental quarks. It is renormalized around the quark mass $\alpha_{\rm DC}= 6\pi/\left[(11N_{\rm DC}-2N_f)\log{\left(m_{\mathcal{Q}}/\Lambda\right)} \right]$. $\zeta$ accounts for group factors and non-perturbative effects. Around the freeze-out temperature, one has $\zeta\simeq 0.25/N_f$, see App.~\ref{sec:quark_FO}. Except if stated otherwise, we assume a single quark flavour $N_f = 1$.

\section{QUARK SQUEEZE-OUT}
\label{sec:quark_SO}
\subsection{Baryon survival fraction}
A single quark cannot pass through the wall boundary. As the quark approaches the wall, it feels a strong force pulling it back within the deconfined phase because of its dark color charge. One may imagine a flux tube between the quark and the boundary, which in the absence of light quarks which could nucleate and break the tube, quickly springs the quark back into the pocket. 
 Heavy quarks are vastly separated by $d\simeq \Lambda^{-1}(\alpha_{\rm DC}^2M_{\rm pl}/m_{\mathcal{Q}})^{1/3} \gg \Lambda^{-1}$, see Eq.~\eqref{eq:Yqinf}. The probability for $k$ quarks to group into a color-singlet configuration which could enter bubble walls is suppressed by $(d\Lambda)^{-3k}\ll 1$, see Eq.~\eqref{eq:poisson_dist}. Hence, heavy quarks are squeezed into contracting pockets of the deconfined phase until they annihilate with anti-quarks.
During this process, a fraction of quarks is successful at forming baryons due to three contributions
\begin{equation}
\label{eq:S_fac}
    S \equiv \frac{3N_{3}^f}{N_1^i}= S_{\rm asym}+ S_{\rm bulk} +S_{\rm bdy},
\end{equation}
where $N_i$ is the total number of particles in a given pocket, $i=1,\cdots N_{\rm DC}$ being its quark number and $i=0$ denoting gluons. The superscripts $i/f$ refers to before/after the confining phase transition. We refer to Eq.~\eqref{eq:quark_abundance_initial} in App.~\ref{app:Quark_SO} for the expression of the initial quark number $N_1^i$.
The first piece $S_{\rm asym} = 1/\sqrt{2N_1^i}$ in Eq.~\eqref{eq:S_fac} is the accidental asymmetric fraction resulting from the relative number of quark minus anti-quark being a stochastic variable following Poisson statistics, see App.~\ref{app:Quark_SO}. The second piece $S_{\rm bulk}$ is the fraction of quarks successful at forming baryons in the bulk which later escape through the boundary \cite{Asadi:2021pwo}. The third piece $S_{\rm bdy}$, which is one of the novelty of the present work, is the fraction of quarks successful at forming baryons directly at the wall boundary (`bdy').\\
\subsection{Boltzmann equations}
In order to calculate the symmetric component of the quark survival fraction
\begin{equation}
\label{eq:S_sym_def}
    S_{\rm sym} = S_{\rm bulk}+S_{\rm bdy} = \frac{3\int dN_{3}}{N_1^i},
\end{equation}
it is necessary to keep track of the evolution of the different particle number $N_i$ in the pocket of volume $V$. The rate of change in particles number obey to the following set of Boltzmann equations
\begin{equation}
\label{eq:L_i_C_i}
   \frac{d N_i}{dt} = \Gamma_{\rm bulk}[i]+\Gamma_{\rm bdy}[i],
\end{equation}
where the two interaction rates per pocket respectively account for processes occurring in the bulk and at the boundary. \\

\begin{figure*}[ht!]
    \centering
    \includegraphics[width=0.7\textwidth]{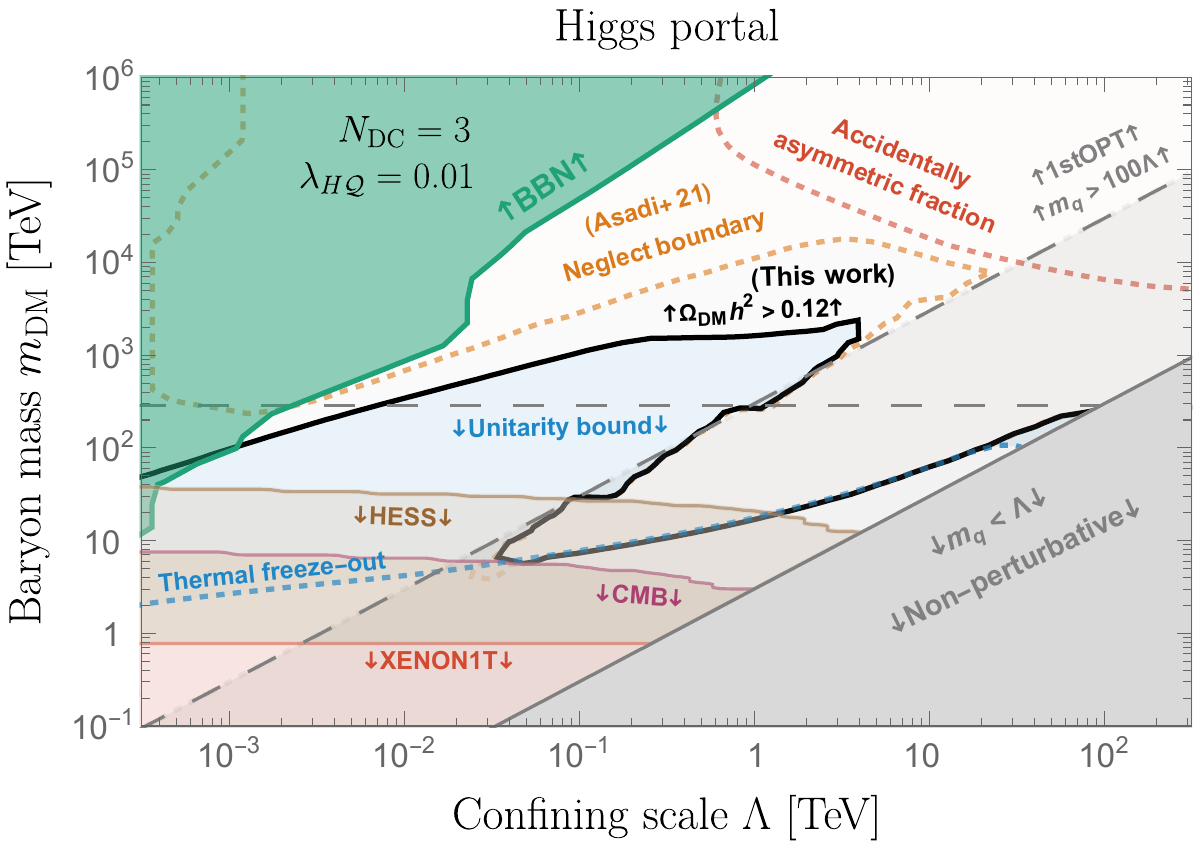}
    \caption{Dark baryon relic abundance including the second quark annihilation stage in the pocket. We numerically solve the Boltzmann equations including processes occurring both in the bulk and at the boundary (\textbf{black line}). To explain DM, dark baryon can have much larger masses than with the thermal freeze-out scenario (\textbf{dotted blue}). The usual unitarity bound on thermal DM \cite{Griest:1989wd} (\textbf{dashed blue}) is naturally evaded.   Previous works \cite{Asadi:2021pwo} neglected effects from the boundary (\textbf{dotted orange}). Quarks and anti-quark can not be depleted below the residual asymmetric fraction accidentally generated in each pocket (\textbf{dotted red}). For $m_{\mathcal{Q}} \lesssim \Lambda$, the $SU(N_{\rm DC})$ coupling constant binding quarks together becomes strong (\textbf{gray region}). For $m_{\mathcal{Q}} \gtrsim 100\Lambda$, the PT is first-order (\textbf{dashed gray}).    We assume that the dark $SU(N_{\rm DC})$ communicate with the SM through Higgs mixing $\lambda_{H\mathcal{Q}}\overline{H}H\overline{\mathcal{Q}}\mathcal{Q}/m_{\mathcal{Q}}$, see App.~\ref{app:heavY_quark_portal}. The glueball can be long-lived leading to BBN constraints \cite{Kawasaki:2017bqm} (\textbf{green}). Baryon annihilation in the galaxy leads to cosmic-rays fluxes constrained by HESS \cite{Profumo:2017obk} (\textbf{brown}), baryon annihilation after recombination leads to CMB constraints \cite{Profumo:2017obk} (\textbf{purple}), direct interaction with nucleons leads to direct detection (DD) constraints from XENON1T \cite{XENON:2023cxc} (\textbf{red}). We fix $\lambda_{H\mathcal{Q}}=0.01$, knowing that smaller $\lambda_{H\mathcal{Q}}$ would lead to stronger BBN constraints and weaker DD constraints. See App.~\ref{app:pheno} for more details on the phenomenology.}
    \label{fig:mDM_Lambda_EW_portal}
\end{figure*}

\begin{figure*}[ht!]
    \centering
    \includegraphics[width=0.7\textwidth]{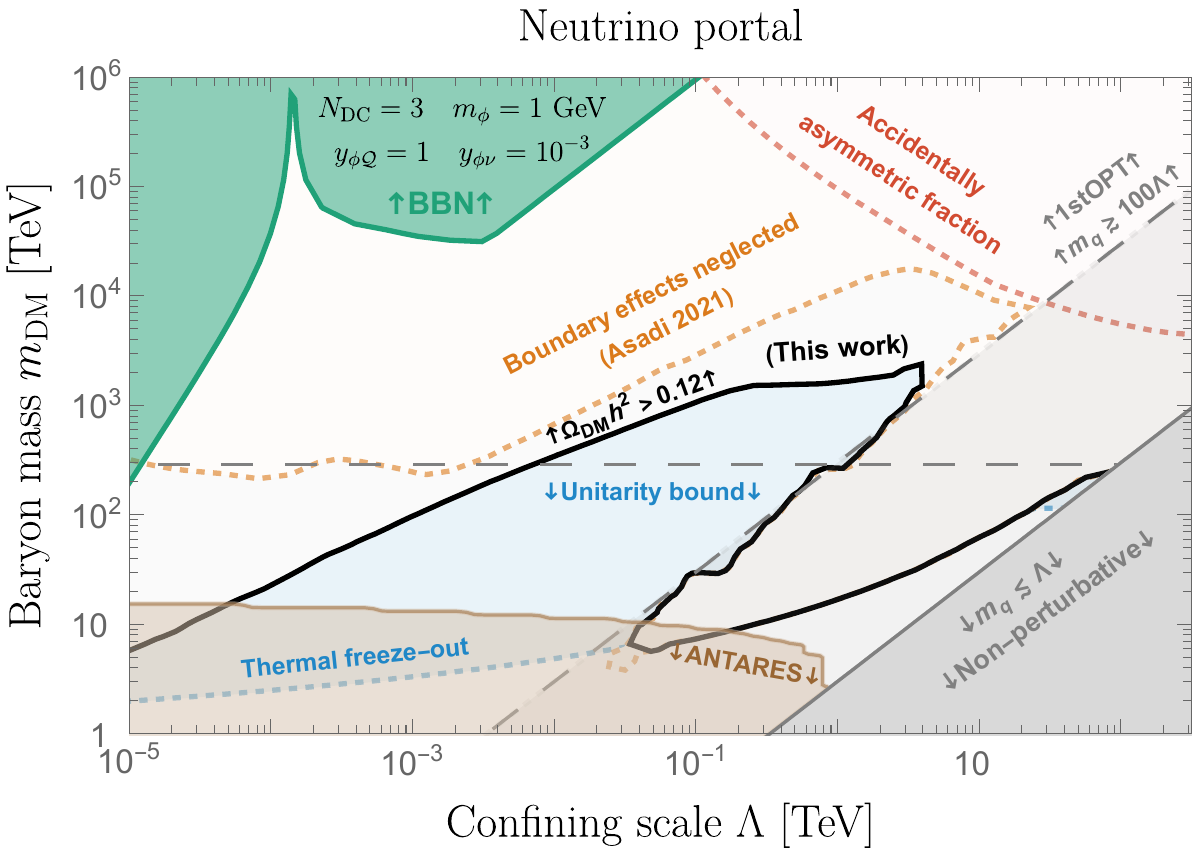}
    \caption{ Same as Fig.~\ref{fig:mDM_Lambda_EW_portal} where we assume that the dark $SU(N_{\rm DC})$ communicates with the SM through a light scalar with Yukawa coupling with heavy dark quarks and SM neutrino. This allows for a longer glueball lifetime, resulting in a reduced impact from both entropy injection and BBN constraints. We fix the light scalar mass to $m_\phi = 1~\rm GeV$, Yukawa couplings to heavy quarks $y_{\phi\mathcal{Q}}=1$ and neutrinos $y_{\phi\nu} = 10^{-3}$, a ballpark away from current constraints \cite{Berryman:2022hds} which minimizes the BBN constraint from glueball decay (\textbf{green}). See App.~\ref{sec:light_scalar_portal} for more details.}
    \label{fig:mDM_Lambda_scalar_portal}
\end{figure*}
\subsection{Bulk effects}
In the bulk, pairs of quark can combine to form a di-quark which can combine with another quark to form a tri-quark, and so one until a baryon $\mathcal{B}$ made of $N_{\rm DC}$ is formed. 
The evolution of the number of each particles in the pocket follows the set of Boltzmann equations
   \begin{equation}
  \label{eq:finiteblz}
   \Gamma_{\rm bulk}[i]=-  \sum_{\rm a+b = c+d}s_{\rm ab,cd}^i \frac{\langle\sigma v\rangle_{abcd}^{\rm bulk}}{V}\left(N_aN_b-N_c N_d f_{abcd}\right),
   \end{equation}
   where $s_{\rm ab,cd}^i$ is the net number of $i$ particles destroyed in the process $ab \to cd$ with cross-section $\left<\sigma v\right>_{\rm ab cd}$ and $f_{abcd}\equiv n^{\rm eq}_a n^{\rm eq}_b/(n^{\rm eq}_c n^{\rm eq}_d)$ with $n_x^{\rm eq}$ the number densities $n_x=N_x/V$ at equilibrium. Bulk terms in Eq.~\eqref{eq:finiteblz} have been extensively studied in \cite{Asadi:2021pwo} which we refer for more details, together with App.~\ref{app:Quark_SO}. We have successfully replicated their results to an excellent degree.
   In the next section, we discuss the boundary term $\Gamma_{\rm bdy}[i]$, which were not accounted for in prior works \cite{Asadi:2021pwo, Asadi:2021yml}. \\
\subsection{Boundary Effects}
A single quark cannot escape from the pocket. However, multiple quarks can simultaneously attempt to depart from the pocket by converging to the same location on the wall in a color-neutral configuration, such as a meson or a baryon. In this case, the wall would not pull back on the quarks allowing them to escape as a hadron.
The boundary term in the Boltzmann equation can be decomposed as a sum of filtering rates
   \begin{align}
       \Gamma_{\rm bdy}[1] &= -\Gamma_{(1,-1) \rightarrow  \mathcal{M} }^{\rm bdy}-3\Gamma_{(1,1,1) \rightarrow 3}^{\rm bdy}-\Gamma_{(1,2) \rightarrow 3}^{\rm bdy},\notag\\
       \Gamma_{\rm bdy}[2] &= -\Gamma_{(1,2) \rightarrow 3}^{\rm bdy},\\
       \Gamma_{\rm bdy}[3] &= -\Gamma_{3 \rightarrow 3}^{\rm esc},\notag
   \end{align}
where $\Gamma_{(a,b) \rightarrow (c,d) }^{\rm bdy}$ is the number of processes $ab\rightarrow cd$ occurring at the wall boundary per unit of time. Here $\mathcal{M}$ denotes  mesons. The baryon escape term $\Gamma_{3 \rightarrow 3}^{\rm bdy}$ is renamed $\Gamma_{3 \rightarrow 3}^{\rm esc}$ to differentiate it from other boundary terms where bound-states actually form at the boundary. Instead, $\Gamma_{3 \rightarrow 3}^{\rm esc}$ is the escape rate of baryons formed in the `bulk'. In contrast to the other boundary terms, the escape term was already included in prior works \cite{Asadi:2021pwo, Asadi:2021yml}.
Assuming that the particles are interacting weakly in the bulk, then the probability of $k$ particles $i$ are meeting through the wall at the same time is given by the Poisson distribution:
\begin{equation}
\label{eq:poisson_dist}
   P_{\lambda_i}\left(k\right)=\frac{\lambda_i^k}{k!}e^{-\lambda_i}\;.
   \end{equation}
The quantity $\lambda_i$ is the mean number of particles $i$ passing through a slit of surface area $\sigma = f_\sigma \pi\ell^2$ during $\Delta t = \ell/v_{\mathcal{Q}}$ where $\ell \simeq 1/\Lambda$ is the confining length and $v_{\mathcal{Q}}$ is the particle velocity in the plasma frame which we approximate to be equal for all species. We have introduced $f_\sigma$ a form factor encoding uncertainties on the cross-section, which we set to $f_\sigma = 1$ in the main text and explore its impact on the results in App.~\ref{app:Quark_SO}. Denoting by $\bar{v}_\perp = v_{\mathcal{Q}}/4+v_w$ the average of the normal component of the particle velocity in the wall frame, $\bar{v}_\perp\, n_i$ is the flux through the wall and we obtain:
\begin{equation}
\label{eq:lambda_i_def}
\lambda_i = \Delta t \,\bar{v}_\perp n_i \,\sigma \simeq f_\sigma \pi n_i/4\Lambda^3,
\end{equation}
Denoting by $R$ the radius of the pocket, we deduce the baryon formation rate out of 3 incoming quarks: 
\begin{align}
\label{eq:gamma111}
\Gamma_{(1,1,1) \rightarrow 3}^{\rm bdy}&=\oint_{\partial V}\frac{dS}{\sigma\Delta t}\sum_{k=3}^\infty P_{\lambda_1}\left(k\right) \simeq v_{\mathcal{Q}}\frac{9 f_\sigma^2 }{2048 \Lambda^6} \frac{N_1^3}{R^7}\;,
\end{align}
where the sum include the possibility to have more quarks than necessary to form a bound-state. 
We also consider the meson production from a quark/anti-quark pair:
\begin{align}
\label{eq:gamma11}
    \Gamma_{(1,-1) \rightarrow \mathcal{M}}^{\rm bdy}\hspace{-0.1cm}&=\oint_{\partial V}\hspace{-0.05cm}\frac{dS}{\sigma\Delta t} \left[\sum_{k=1}^\infty P_{\lambda_1}\hspace{-0.1cm}\left(k\right) \right]^2 \hspace{-0.2cm}\simeq  v_{\mathcal{Q}}\frac{9f_\sigma}{64 \Lambda^3}\frac{N_1^2}{R^4} ,
\end{align}
and the baryon production out of a quark / di-quark pair:
\begin{align}
\label{eq:gamma12}
    \Gamma_{(1,2) \rightarrow 3}^{\rm bdy}&=\oint_{\partial V}\frac{dS}{\sigma\Delta t} \left[\sum_{k=1}^\infty P_{\lambda_1}\left(k\right) \right]\left[\sum_{k=1}^\infty P_{\lambda_2}\left(k\right) \right]\notag \\
    &\simeq v_{\mathcal{Q}}\frac{9f_\sigma}{64 \Lambda^3}\frac{N_1N_2}{R^4}\;.
\end{align}
When already formed in the bulk through Eq.~\eqref{eq:finiteblz}, baryon can freely escape (`esc') the pocket with a rate:
\begin{align}
    \Gamma_{3 \rightarrow 3}^{\rm esc} &= \oint_{\partial V} \hspace{-0.05cm}\frac{dS}{\sigma \Delta t} \left[\sum_{k=1}^\infty k P_{\lambda_3}\left(k\right) \right]= \frac{3v_{\mathcal{Q}} N_3}{4R}.
\end{align}
We are now able to express the quark survival factor in Eq.~\eqref{eq:S_sym_def} which receives contribution from both bulk and boundary terms:
\begin{equation}
\label{eq:S_symm_def}
S_{\rm sym}=\int_{R_i}^0 \frac{dR}{v_w}\left( \dot{S}_{\rm bulk}+\dot{S}_{\rm bdy}\right),
\end{equation}
where $R_i$ and $v_w$ are given in Eqs.~\eqref{eq:Ri} and \eqref{eq:v_w_contraction}, and:
\begin{align}
    &\dot{S}_{\rm bulk}=\Gamma_{3 \rightarrow 3}^{\rm esc.}/N_1^i\\
    &\dot{S}_{\rm bdy}=(3\Gamma_{(1,1,1) \rightarrow 3}^{\rm bdy}+\Gamma_{(1,2)\rightarrow 3}^{\rm bdy})/N_1^i.
\end{align}
In prior works \cite{Asadi:2021pwo, Asadi:2021yml}, only the `bulk' term $\dot{S}_{\rm bulk}$ has been included. The surviving abundance is shown in Fig.~\ref{fig:fractional_abundance} after numerically integrating the set of coupled Boltzmann equations in Eq.~\eqref{eq:L_i_C_i}. We refer to App.~\ref{app:Quark_SO} for more details.

\section{BARYON DILUTE-OUT}
\label{sec:quark_DO}
According to Ref.~\cite{Asadi:2022vkc}, 3-to-2 interactions are efficient enough to maintain glueballs at thermal equilibrium with themselves right after the end of the PT as long as $\Lambda \lesssim 10^5~\rm TeV$. Then, 3-to-2 interaction freeze-out around the temperature $x_{\rm FO}= m_{\rm GB}/T_{\rm FO} \sim 20$, which, the glueball mass being $m_{\rm GB}\simeq 7\Lambda$ \cite{Mathieu:2008me}, reduces to $T_{\rm FO}\sim 0.3T_c$. The frozen-out glueball relic abundance is \cite{Forestell:2016qhc}:
\begin{equation}
\label{eq:YGB_app}
Y_{\rm GB} = R_{\rm GB} /x_{\rm FO},\qquad R_{\rm GB}\simeq 2\times 10^{-4},
\end{equation}
where $R_{\rm GB}\equiv s_{\rm GB}/s_{\rm SM}$ is the glueball-to-SM entropy ratio, here set to its equilibrium value, see Eq.~\eqref{app:quark_DO}, which as stated above, is a valid approximation as long as $\Lambda \lesssim 10^5~\rm TeV$.
Glueballs decay into SM around the temperature:
\begin{equation}
\label{eq:Tdec_app}
T_{\rm dec}\simeq 1.3g_*^{1/4}\sqrt{\Gamma_{\rm GB} M_{\rm pl}},
\end{equation}
where $\Gamma_{\rm GB}$ is the glueball decay rate. The later is generally suppressed by large power of $\Lambda/m_{\mathcal{Q}}$ such that glueballs can be long-lived and dominate the energy density of the universe below the temperature:
\begin{equation}
\label{eq:Tdom_app}
T_{\rm dom} = \frac{4}{3}m_{\rm GB}Y_{\rm GB},
\end{equation}
if $T_{\rm dom}> T_{\rm dec}$. The glueball decay increases the SM entropy by the ratio, see App.~\ref{app:quark_DO}:
\begin{equation}
\label{eq:D_fac}
D \equiv \frac{s_{\rm SM}^f}{s_{\rm SM}^i}  \simeq 1+ \frac{T_{\rm dom}}{T_{\rm dec}},
\end{equation}
where $i/f$ refers to before and after the decay.
As a results, the residual dark baryon population is diluted and the DM relic abundance today reads
\begin{equation}
\label{eq:Y_DM_final_formula}
{Y_{\rm DM}^0~ \simeq~ \frac{S}{D}\, \frac{2Y_{\mathcal{Q}}^{\rm FO}}{N_{\rm DC}}}.
\end{equation}
 The expression in Eq.~\eqref{eq:Y_DM_final_formula} accounts for quarks freeze-out through $Y_{\mathcal{Q}}^{\rm FO}$ defined in Eq.~\eqref{eq:Yqinf} the factor of $2$ including the contribution from anti-quarks, quarks squeeze-out through the survival factor $S\leq 1$ defined in Eq.~\eqref{eq:S_fac}, and baryons dilution through $D\geq 1$ defined in Eq.~\eqref{eq:D_fac}.

\section{RESULTS}
\label{sec:pheno}
\subsection{Dark Matter abundance}
\textbf{Numerical treatment.}
We calculate the dark baryon abundance defined in Eq.~\eqref{eq:Y_DM_final_formula} by solving the system of coupled Boltzmann equations in Eq.~\eqref{eq:L_i_C_i}. The time evolution of the quark abundance inside and outside a typical pocket of deconfined phase is shown in Fig.~\ref{fig:fractional_abundance}.
The dark baryon mass $m_{\rm DM}$ and confining scale $\Lambda$ which reproduce the observed DM relic abundance $\Omega_{\rm DM}h^2\simeq 0.12$ are shown indicated with the black lines in Figs.~\ref{fig:mDM_Lambda_EW_portal} and \ref{fig:mDM_Lambda_scalar_portal}. The second quark annihilation stage due to squeezing predicts a DM mass up to $m_{\rm DM} \sim 10^3~\rm TeV$, two orders magnitude above the thermal freeze-out prediction shown in dotted blue line, and about one order of magnitude above the unitarity bound shown in dashed blue line \cite{Griest:1989wd}. Neglecting for the production of dark baryons at the pocket boundary would overestimate the DM mass by one order of magnitude, see dotted orange line in Figs.~\ref{fig:mDM_Lambda_EW_portal} and \ref{fig:mDM_Lambda_scalar_portal}. We find that entropy injection following glueball decay only substantially dilutes the DM abundance in regions where the glueball is so long-lived that it is already ruled out by BBN. For instance, this can be seen in the curvature of the orange dashed line in the top-left corner of Fig.~\ref{fig:mDM_Lambda_EW_portal}. \\
\textbf{Analytical treatment.}
Analytical expressions for the dark baryon abundance are calculated in App.~\ref{app:Quark_SO} and plotted against the numerical results in Fig.~\ref{fig:ana_vs_num}. \\
\textbf{Source of uncertainties.}
We identify two main sources of uncertainties in our treatment. At first, the pocket wall velocity derived in Eq.~\eqref{eq:v_w_contraction} can be understood as an upper bound $v_{\rm max}$ which is reached if the latent heat is efficiently released to the whole universe and not only to the region close to the wall. Also it does not account for the effects from heavy quarks on the friction pressure \cite{Asadi:2021pwo}, whose precise derivation is left for future studies. In Fig.~\ref{fig:mDM_Lambda_vary}, we tune down the maximal wall velocity $v_{\rm max}$ in Eq.~\eqref{eq:v_w_contraction} by factors $10^{-2}$ and $10^{-4}$. For $v_w \simeq 10^{-4}v_{\rm max}$, we find that boundary effects disappear completely. Instead, bulk effects disappear in the ballpark $\Lambda \in [1~\rm TeV,10~\rm TeV]$ and $m_{\rm DM}\simeq 10^4~\rm TeV$ where the dark baryon abundance is then given by the accidental asymmetry, in agreement with \cite{Asadi:2021yml}. The second source of uncertainties is the value of the cross-section $\sigma = f_\sigma \pi/\Lambda^2$ for bound-state formation at the boundary which enters Eq.~\eqref{eq:lambda_i_def}. In Fig.~\ref{fig:mDM_Lambda_vary}, we scan over the values $f_\sigma \in [0.1,10]$, showing that the predicted DM mass can vary by about one order of magnitude around its mean value.

\subsection{Phenomenology}
We consider two benchmark portal to the SM: Higgs portal and neutrino portal.\\
\textbf{Higgs portal.}
We suppose heavy quarks receive their mass from $\phi\overline{\mathcal{Q}}\mathcal{Q}$ where $\phi$ is a singlet scalar getting vacuum expectation value. We assume that $\phi$ mixes with Higgs $\lambda_{H\phi} H\phi$. This generates the dimension 6 operator $\mathcal{O}_6=\overline{H}H\overline{\mathcal{Q}}\mathcal{Q}$ with coefficient $c_6=\lambda_{H\mathcal{Q}}/m_{\mathcal{Q}}$ and $\lambda_{H\mathcal{Q}}=\sqrt{2}\lambda_{h\phi}(m_\mathcal{Q}/m_\phi)^2$. This allows the glueball to decay into SM and gives constraints from indirect detection (ID), Cosmic Microwave Background (CMB), direct detection (DD) and Big-Bang Nucleosynthesis (BBN), which we show in Fig.~\ref{fig:mDM_Lambda_EW_portal}. Large $\lambda_{H\mathcal{Q}}$ gives lower BBN constraints but larger DD constraints, while ID and CMB are independent of $\lambda_{H\mathcal{Q}}$. We fix $\lambda_{H\mathcal{Q}}=0.01$ in Fig.~\ref{fig:mDM_Lambda_EW_portal}. We refer to App.~\ref{app:heavY_quark_portal} and \ref{app:pheno} for more details on the Higgs portal and the associated phenomenology.\\
\textbf{Neutrino portal.} We now consider the possibility that the scalar $\phi$, even though it interacts with heavy quarks through $\phi\overline{\mathcal{Q}}\mathcal{Q}$, does not actually give mass to these heavy quarks and is significantly lighter. We suppose that mediation with SM takes place via SM neutrino $\phi \overline{\nu}_L\nu_L$. The BBN constraints are weaker because the lighter mediator allows a shorter glueball lifetime but also because of the elusive nature of neutrino. For the same reason, the portal secludes the dark baryon DM from direct detection, CMB and collider constraints. Only indirect-detection constraints inferred from neutrino telescopes like ANTARES \cite{Albert:2016emp} can be applied with a region constrained around $m_{\rm DM}\sim 10~\rm TeV$, see Fig.~\ref{fig:mDM_Lambda_scalar_portal}. We refer to App.~\ref{sec:light_scalar_portal} for more details on the neutrino portal and to App.~\ref{app:pheno} for its phenomenology.\\
\textbf{Gravitational waves.}
A well-studied observable from first-order phase transition is gravitational waves (GW) from bubble collision, sound waves and turbulence \cite{Witten:1984rs,Caprini:2015zlo,Schwaller:2015tja,Gouttenoire:2022gwi}. The would-be GW energy density today, here expected to be dominantly sourced by sound waves \cite{Gouttenoire:2021kjv}, is $\Omega_{\rm GW}h^2 \simeq 10^{-6} v_w^2 (R_0 H)^2 \alpha^2$ \cite{Gouttenoire:2022gwi} with $\alpha\equiv \LL/\rho_{\rm SM}\simeq 3(\LL/T_c^4)/g_{*,\rm SM}$ being the latent heat fraction.
It was already known \cite{Huang:2020crf, Halverson:2020xpg, Reichert:2021cvs,Ares:2021nap,Morgante:2022zvc,Pasechnik:2023hwv} that GW from Yang-Mills phase transition were challenging to observe with future interferometers due to the small bubble size at percolation $R_0H \sim 10^{-5}$, see Eq.~\eqref{eq:v_w_Hubble} and the moderate latent heat fraction $\alpha \sim 0.1$ ($N_{\rm DC}=3$, $g_{*,\rm SM}\sim 40$). The results concerning the small bubble wall velocity $v_w\sim 10^{-6}$ presented here in Eq.~\ref{eq:v_w_Hubble} decisively rule out any possibility to observe GW from Yang-Mills confinement in the foreseeable future. Alternatively, the presence of a matter era due to the possibility for glueballs to dominate the energy budget of the universe before decaying could leave signatures in the spectrum from pre-existing gravitational waves, e.g. from a network of cosmic strings or inflation \cite{Giovannini:1998bp,Nakayama:2008ip,Cui:2017ufi,Cui:2018rwi,Gouttenoire:2019kij,Gouttenoire:2019rtn,Gouttenoire:2021jhk,Ghoshal:2023sfa}.

\section{CONCLUSION}
We have studied confinement of $SU(N_{\rm DC})$ with $N_{\rm DC}=3$ in presence of quarks with mass much larger than the confining scale $m_{\mathcal{Q}} \gg \Lambda$. The confining phase transition is first-order \cite{Saito:2011fs}. Due to their large separation after freeze-out, quarks can not penetrate inside bubble walls \cite{Asadi:2021pwo, Asadi:2021yml}.
In the absence of light quarks in the spectrum, the only way for quarks to enter bubble walls is by forming color-neutral bound states in the bulk or directly at the wall boundary. 
In this work, we estimate how the baryon relic abundance is impacted by bound-state formation at the boundary, an effect which was overlooked in previous studies \cite{Asadi:2021pwo, Asadi:2021yml}. We find that $SU(3)$ dark baryon can be DM with a mass around $m_{\rm DM} \sim 10^3~\rm TeV$, one order of magnitude above the unitarity bound \cite{Griest:1989wd} instead of two has previously thought \cite{Asadi:2021pwo, Asadi:2021yml}. Introducing Higgs and neutrino portals, we found that due to the large DM mass, this scenario is hardily testable at telescopes, colliders and by cosmological probes.
Finally, Yang-Mills confinement proceeds with the liberation of a considerable amount of latent heat, despite the amount of supercooling being ridiculously small. As a consequence, the bubble wall velocity is particularly low. It is set by the Hubble expansion rate at $v_w \sim 10^{-6}$, much smaller than previous estimates \cite{Huang:2020crf, Halverson:2020xpg, Reichert:2021cvs,Ares:2021nap,Morgante:2022zvc,Pasechnik:2023hwv}, ruling out any prospects for observing gravitational waves arising from bubble dynamics.  The dark baryon DM relic abundance established in the initial studies \cite{Asadi:2021pwo, Asadi:2021yml} remains valid provided that additional factors further reduce the pocket wall velocity by approximately $10^{-4}$ relative to the maximum value presented in Eq.~\eqref{eq:v_w_contraction}, as seen in Fig.~\ref{fig:mDM_Lambda_vary}. The possibility that the wall velocity is much smaller than its maximum value, due to additional effects not accounted in this work like the quark pressure, is left for future studies.  In any case, the studies in Refs.~\cite{Asadi:2021pwo, Asadi:2021yml} limited their scope to $\Lambda \in [\rm TeV,~10~\rm TeV]$, whereas our calculations of the DM relic abundance extend down to confinement scale as low as allowed by BBN which under the assumption of efficient glueballs decay channels means $\Lambda \gtrsim 5~\rm MeV$, see App.~\ref{app:latent_heat}. Finally, we draw attention to the possibility, arising when initial baryonic asymmetry is present, for quarks to be squeezed into quark nuggets \cite{Witten:1984rs,Frieman:1990nh,Zhitnitsky:2002qa,Oaknin:2003uv,Atreya:2014sca,Gresham:2017cvl,Bai:2018vik,Bai:2018dxf,Ge:2019voa}, dark stars, or primordial black holes \cite{Gross:2021qgx}. The potential impact of the boundary effects introduced in this work on such scenarios, especially in the heavy quark limit \cite{Gross:2021qgx}, is a topic left for future studies.

{\bf Acknowledgements.}---%
The authors thank Pouya Asadi, Iason Baldes, Xucheng Gan, Guy D. Moore, Filippo Sala and Juri Smirnov for valuable conversations.
YG is grateful to the Azrieli Foundation for the award of an Azrieli Fellowship.  EK is supported by the Israel Science Foundation (grant No. 1111/17) and by the Binational Science Foundation  (grants No. 2016153). DL acknowledges funding from the French Programme d’investissements d’avenir through the Enigmass Labex.

\appendix

\clearpage
\appendix
\onecolumngrid

\renewcommand{\tocname}{\large  Table of contents}
 {
 \hypersetup{linkcolor=black}
 \tableofcontents
 }

\fontsize{11}{13}\selectfont

\titleformat{\section}
{\normalfont\fontsize{12}{14}\bfseries  \centering }{\thesection.}{1em}{}
\titleformat{\subsection}
{\normalfont\fontsize{12}{14}\bfseries \centering}{\thesubsection.}{1em}{}
\titleformat{\subsubsection}
{\normalfont\fontsize{12}{14}\bfseries \centering}{\thesubsubsection)}{1em}{}

\titleformat{\paragraph}
{\normalfont\fontsize{12}{14}\bfseries  }{\thesection:}{1em}{}

\section{Dynamics of Yang-Mills confinement}

\subsection{Thermodynamics}
\label{sec:thermo}
We consider a $SU(N)$ gauge theories with $N_f$ favours of quarks:
\begin{equation}
    \mathcal{L} = \mathcal{L}_{\rm SM} - \frac{1}{4}G_{\mu\nu}^a G^{\mu\nu a} + \bar{q}(i\slashed{D} - m_{\mathcal{Q}})q,
\end{equation}
with $D_\mu = \partial_\mu -ig_{\rm D}A_\mu$. The  gauge coupling $\alpha_{\rm DC}=g_{\rm D}^2/4\pi$ runs with the energy scale, here set to the quark mass $\mu=m_{\mathcal{Q}}$, as \cite{ParticleDataGroup:2022pth}:
\begin{equation}
\label{eq:alpha_D_rge}
\alpha_{\rm DC} = \frac{6\pi}{(11N_{\rm DC}-2N_f)\log{(m_{\mathcal{Q}}/\Lambda)}}.
\end{equation}
We examine the scenario where the quarks are heavy $m_{\mathcal{Q}} \gtrsim 100\Lambda$. In this regime, quark confinement can be accurately modeled using an $SU(N)$ Yang-Mills theory with $N_f=0$. The case $N_f = 0$ is known to result in a first-order phase transition and has been extensively studied through lattice simulations \cite{Lucini:2003zr, Lucini:2005vg, Panero:2009tv, Borsanyi:2012ve}. 
A phase transition is said of the first-order if the first derivative of the free energy density $f$, namely the entropy density $s=-\frac{\partial f}{\partial T}$, is discontinuous at some critical temperature $T_c$:
\begin{equation}
\Delta s = -\frac{\partial \Delta f}{\partial T} = \frac{\LL}{T_c},
\end{equation}
where $\LL$ is the latent heat of the phase transition. Using that the free energy $f = \rho-Ts$ is continuous at $T_c$, we get the relation $\Delta \rho =T_c\Delta s=\LL$, such that the latent heat $\LL$ can be interpreted as a gap in energy density. The thermodynamics of 1stOPT can be found in textbooks \cite{Landau:1980mil} and reviews \cite{Landau:1980mil,Fuller:1987ue,Asadi:2021pwo}.
Thermal tunneling in the thin-wall limit is described by the $O_3$-symmetrical euclidean action which can be approximated by:
\begin{equation}
S_3 = 4\pi \sigma R_c^3+\frac{4\pi}{3}R_c^3\Delta f,
\end{equation}
where $\sigma$ is the energy per unit of surface area of the bubble wall, also known as surface tension. Bubbles are nucleated around a radius $R_c$, saddle-point of the bounce action:
\begin{equation}
\frac{S_3}{\partial R_c}=0\quad \implies \quad R_c = \frac{2\sigma}{\Delta f}.
\end{equation}
The lattice fitting of the latent heat and surface tension for $\rm SU(N)$ YM theories is \cite{Lucini:2005vg}:
\begin{equation}
   \LL\simeq (0.76-0.3/N_{\rm DC}^2)^4N_{\rm DC}^2T_c^4,\qquad \textrm{and} \qquad \sigma \simeq (0.015N_{\rm DC}^2-0.1)T_c^3. 
\end{equation}
More precisely, for $N_{\rm DC}=3$ one finds (table 7 and 11 in \cite{Lucini:2005vg} ):
\begin{align}
\label{eq:L_sigma}
\LL=1.413(55)T_c^4,\qquad \textrm{and} \qquad \sigma = 0.0200(6)T_c^3.
\end{align}
In the regime of small supercooling $T_c-T\ll T_c$, which will be justified afterward, the free energy density close to $T_c$ can be Taylor-expanded at first-order:
\begin{equation}
\label{eq:Delta_F_Taylor}
\Delta f (T) \simeq \frac{T_c-T}{T_c}\LL.
\end{equation}
The bounce action at the critical radius can then be written:
\begin{equation}
S_3 =  \frac{16\pi}{3} \frac{\sigma^3}{\Delta f^3} = \frac{16\pi}{3}\left( \frac{\sigma}{T_c^3} \right)^3 \left( \frac{T_c^4}{\LL} \right)^{2} \frac{T_c^3}{(T_c-T)^2}.
\end{equation}
The tunneling rate for thermal phase transition reads \cite{Linde:1981zj}:
\begin{align}
\label{eq:Gamma_tunneling}
&\Gamma \simeq T_c^4\left( \frac{S_3/T}{2\pi} \right)^{3/2}\,\exp\left({-S_3/T_c}\right) \simeq  T_c^4 \exp\left({-\kappa\frac{ T_c^2}{(T_c-T)^2}}\right), 
\end{align}
with:
\begin{equation}
\kappa = \frac{16\pi}{3}\left( \frac{\sigma}{T_c^3} \right)^3 \left( \frac{T_c^4}{\LL}\right)^{2} \simeq 6.7 \times 10^{-5}.
\end{equation}
where we have plugged $SU(3)$ lattice results in Eq.~\eqref{eq:L_sigma}. Bubble nucleation becomes efficient when the tunneling rate of a Hubble volume becomes comparable to the Hubble expansion rate $\Gamma \simeq H^4$. We deduce the nucleation temperature
\begin{equation}
\label{eq:T_c_-_Tn}
    \frac{T_c-T_n}{T_c} \simeq \sqrt{\frac{\kappa}{l}} \simeq 7 \times 10^{-4},
    \end{equation}
  with $l \simeq 10^{2}$.
The very little amount of supercooling ${T_c-T_n}\ll {T_c}$ justifies the validity of the Taylor expansion of the free energy in Eq.~\eqref{eq:Delta_F_Taylor}. 

\subsection{Bubble growth}
\label{app:bubble_growth}

\textbf{Obstruction to bubble expansion.}\\
After nucleation, bubble walls expand under the effect of the latent heat pressure, see Eqs.~\eqref{eq:Delta_F_Taylor} and \eqref{eq:T_c_-_Tn}:
\begin{equation}
\label{eq:Delta_F}
    \Delta f ~=~ \frac{T_c-T}{T_c}\LL ~\simeq ~10^{-3} T_c^4.
\end{equation}
This small value implies that the gluons and glueball phases have almost identical free energies $f=\rho - T s$. However the large latent heat $\LL=1.413(55)T_c^4$ in Eq.~\eqref{eq:L_sigma} suggests that the two phases have very different energy density $\rho$ and entropy density $s$. Indeed the energy and entropy of the gluon phase goes like $\propto \mathcal{O}(N_{\rm DC}^2)$ while the ones of the glueball phase goes like $\propto \mathcal{O}(N_{\rm DC}^0)$. In order to form a glueball pair of mass $2m_{\rm GB}\simeq 14\Lambda$, at least 2 gluons with their $g_G =2(N_{\rm DC}^2-1)$ degrees of freedom each are needed. Assuming their typical energy to be $E_G \sim \Lambda$ and their number density per degree of freedom $n_* = \zeta(3)T_c^3/\pi^2$ (with $T_c \simeq \Lambda$), we find that the gluon-to-glueball conversion process liberates an energy density $\LL \sim (2g_G E_G - 2m_{\rm GB})n_* \sim 2 T_c^4$ for $N_{\rm DC}=3$ not too far from the value predicted by lattice simulations in Eq.~\eqref{eq:L_sigma}. We stress that the previous argument should not be viewed as a rigorous proof but more as a heuristic explanation for why the latent heat is large.  As a consequence, the process of converting gluons into glueballs is very exothermic. If glueballs and gluons are at kinetic equilibrium with the SM, the latent heat released in the glueball phase will be quickly communicated to the gluons phase by SM relativistic particles acting as mediators. In light of the very small amount of the supercooling in Eq.~\eqref{eq:T_c_-_Tn}, if bubble walls move too fast the temperature around the wall would be easily reheated above $T_c$, causing the free energy in Eq.~\eqref{eq:Delta_F} to flip sign and bubble wall to reverse their direction.
We leave the scenario where glueballs and gluons are not at kinetic equilibium with the SM for future studies, the question being whether the glueballs and gluons gas can thermalise through the wall boundary.

\textbf{Bubble wall velocity from Hubble expansion.}
We now assume that a mechanism is at play to exchange heat between the two phase, e.g. kinetic equilibrium of gluons and glueballs with SM, such that the latent heat is quickly released to the surrounding plasma, causing the universe to be continuously reheated as bubbles grow. We introduce the fraction $x$ of the universe in the confined phase ($0\leq x\leq1$). The release of a fraction $dx$ of latent heat increases the universe temperature by:
\begin{equation}
\label{eq:inst_rehe}
dT = \LL \left(\frac{d\rho}{dT}\right)^{-1} dx = \frac{15}{2\pi^2 g_{*}} \frac{\LL }{T^3} dx,
\end{equation}
where $g_{*}= 2(N_{\rm DC}^2-1)(1-x)+g_{\rm SM}$ is the number of relativistic degrees of freedom in the thermal bath. 
The evolution of the temperature is a competition between adiabatic cooling due to Hubble expansion and heating due to latent heat conversion:
\begin{equation}
\label{eq:T_dot}
\dot{T} = -HT +   \frac{\LL T}{4\rho} \dot{x}=  -HT +  \frac{3R^2\LL T}{8R_0\rho} v_w,\qquad \rho \equiv \pi^2g_{*}T^4/30,
\end{equation}
where we have used that $x = (R/R_0)^3/2$ where $R_0$ the bubble radius at percolation defined by $x=1/2$.
An estimate for the wall velocity can be found as resulting from a detailed balance between the rates at which the bubble wall heats and cools the surrounding plasma \cite{Witten:1984rs,Asadi:2021pwo}.
On the one hand, the plasma around the wall boundary is heated due to the latent heat conversion at a rate (for instance due to gluons accumulating in front of the wall):
\begin{equation}
\dot{T}_{\rm heat} \sim \LL\,\Lambda\, \left(\frac{d R}{dt}\right) \times \frac{dT}{d\rho} \sim \Lambda\, \frac{d R}{dt} ,
\end{equation}
where $dR/dt$ is the wall velocity, $\Lambda^{-1}$ is the wall thickness. 
On the other hand, due to the temperature gradient the same plasma cools down at a rate:
\begin{equation}
\dot{T}_{\rm cooling} \sim -K \nabla^2 T \sim \Lambda^{-1} \frac{T_{\rm wall}-T}{\Lambda^2},
\end{equation}
where the transport coefficient $K$ and the gradient length scale $\nabla$ are both given by $\Lambda^{-1}$. 
We set the plasma temperature at the wall at the maximal value $T_{\rm wall}\simeq T_c$. 
The wall velocity results from the balance between the two thermal rates \cite{Witten:1984rs,Asadi:2021pwo}:
\begin{equation}
\dot{T}_{\rm heat} \sim \dot{T}_{\rm cooling} \quad \implies \quad v_w \sim \frac{T - T_c}{T_c}, \label{eq:radius_time}
\end{equation}
where we have replaced $\Lambda \simeq T_c$. The assumption that the wall temperature $T_{\rm wall}\simeq T_c$ and the gradient $\nabla \simeq \Lambda^{-1}$ are set to their largest possible values imply that Eq.~\eqref{eq:radius_time} can be considered as an upper limit for the wall velocity.
Rewriting Eq.~\eqref{eq:radius_time} as $T=T_c(1-v_w)$ and plugging into Eq.~\eqref{eq:T_dot} leads to
\begin{equation}
    \dot{v}_w=H-\frac{3\L R^2 v_w}{8\rho R_0^3}\;,\quad \dot{R}=v_w
\end{equation}
Eliminating the time variable via $dv_w/dR=\dot{v}_w/\dot{R}$, one arrives at
\begin{equation}\label{eq:dvdR}
    \frac{dv_w}{dR}=\frac{H}{v_w}-\frac{3\LL R^2}{8\rho R_0^3}
\end{equation}
The later equation has an attractor solution in which the universe temperature is kept constant:
\begin{equation}
\label{eq:vw_instant_therm}
\dot{T} = 0\quad \implies \quad v_w ~\simeq~ HR_0\frac{8\rho}{3\LL} \left(\frac{R_0}{R}\right)^2 \simeq 100 HR_0\left(\frac{T_c^4}{\LL}\right)\left(\frac{g_*}{100}\right)\left(\frac{R_0}{R}\right)^2,
\end{equation}
Eq.~\eqref{eq:vw_instant_therm} implies that the bubble wall velocity, under the assumption of instantaneous reheating in Eq.~\eqref{eq:inst_rehe}, is controlled by the rate of universe expansion. In principle, other contribution to the friction pressure can slow down the wall, e.g. impact of the heavy quarks. In that sense, Eq.~\eqref{eq:vw_instant_therm} is an upper limit on the wall velocity.

\subsection{Bubble percolation and coalescence}

The evolution of the phase transition can be described by the fraction of the universe in the confined phase \cite{Rubakov:2017xzr}:
\begin{equation}
\label{eq:x_def}
x(t) = \int_{t_c}^td\tilde{t}~\Gamma(\tilde{t})\frac{4\pi}{3}R^3(t,\tilde{t})(1-x(\tilde{t})),
\end{equation}
where $t_c$ is the time at which the transition becomes energetically  allowed, $\Gamma(t)$ is the nucleation per unit of volume at time $t$, defined in Eq.~\eqref{eq:Gamma_tunneling}, and $R(t,\tilde{t})=\int_{t'}^t d\tilde{t}/a(\tilde{t})$ is the radius at time $t$ of a bubble nucleated at time $\tilde{t}$. Taking the time derivative of Eq.~\eqref{eq:x_def} leads to:
\begin{equation}
\dot{x}(t) = \Gamma(t)(1-x(t))\frac{4\pi}{3}R_c^3(t) + \int_{t_c}^t d\tilde{t}~\Gamma(\tilde{t})4\pi R^2(t,\tilde{t})\dot{R}(t,\tilde{t})(1-x(t)),
\end{equation}
where $R_c$ is the bubble radius at nucleation. Let us introduce the temperature difference $\delta \equiv (T_c-T)/T_c$.
Per definition, at $t_c$ we have $\delta(t_c) =0$.
After $t>t_c$, the temperature evolves under Hubble expansion as $T(t) =T_c\,e^{- \int_{t_c}^t d\tilde{t} H}$ which implies
\begin{equation}
\label{eq:delta_H_ttc}
t> t_c:\qquad \delta(t) \simeq H(t-t_c),
\end{equation}
As soon as the temperature reaches $\delta \sim \sqrt{\kappa} \simeq 0.008$ at a time $t_{n}$, the tunneling rate in Eq.~\eqref{eq:Gamma_tunneling} increases very fast, and bubble nucleation becomes very efficient, leading the universe temperature to quickly increase as dictated by Eq.~\eqref{eq:T_dot}:
\begin{equation}
\label{eq:delta(t)_x(t)}
t>t_{n}:\qquad \delta(t)~ \simeq ~\frac{\LL}{4\rho} x(t) .
\end{equation}
This back-react on the tunneling rate and nucleation becomes inefficient. Plugging Eq.~\eqref{eq:T_c_-_Tn} into Eq.~\eqref{eq:delta_H_ttc} implies that the time scale during which nucleation is efficient is:
\begin{equation}
\Delta t ~\simeq ~t_n-t_c \simeq \sqrt{\frac{\kappa}{l}}H^{-1},
\end{equation}
where $l\simeq 10^{2}$.
We deduce the bubble radius at percolation:
\begin{equation}
\label{eq:R_0_def_ana}
R_0 \simeq v_w\Delta t \simeq \delta^2(t_n) H^{-1} \simeq \frac{10^{-7}}{\Lambda}\left( \frac{M_{\rm pl}}{\Lambda} \right),
\end{equation}
where we have used $v_w \simeq \delta(t_n)$ in Eq.~\eqref{eq:radius_time} and  $\delta(t_n) \simeq\sqrt{\kappa/l} \simeq  7\times 10^{-4}$ in Eq.~\eqref{eq:T_c_-_Tn} and $g_* \simeq 100$ in $H$.
The bubble radius a percolation $R_0$ can also be calculated numerically. We introduce the bubble number density $n_{\rm bub}=x/(4\pi R^3/3)$. Its time derivative is set by $\dot{n}_{\rm bub}=\Gamma (1-x)$ where $\Gamma$ is the tunneling rate per unit of volume in Eq.~\eqref{eq:Gamma_tunneling}. Eliminating the time variable via $d/dt=v_w d/dR$, we obtain the first line of:
\begin{align}
\frac{dx}{dR}&=\frac{4\pi R^3\Gamma_{\rm bub}}{3v_w} +\frac{3x}{R}\\
\frac{dv_w}{dR}&=-\frac{H }{v_w}+\frac{{\LL} dx}{4\rho dR}.
\end{align}
The second line comes from Eq.~\eqref{eq:T_dot} with the replacement $\dot{T}=-\dot{v}_w T_c$ and  $d/dt=v_w d/dR$.
Numerically solving the previous system of equations, we find that the spectrum of percolation radii peaks around \cite{Asadi:2021pwo}:
\begin{equation}
\label{eq:R_0_def}
R_0=R\left(x=1/2\right) ~\simeq ~\frac{10^{-6}}{\Lambda}\left( \frac{M_{\rm pl}}{\Lambda} \right)^{0.9},
\end{equation}
in close agreement with the analytical derivation in Eq.~\eqref{eq:R_0_def_ana}.
Due to the decrease in surface area, it is energetically favourable for little bubbles to coalesce into bigger bubbles \cite{Witten:1984rs}. This process takes some finite time $t_{\rm coal}$. When the coalescence time is much faster than the percolation time scale, bubbles merge together to reach the radius \cite{Asadi:2021pwo}:
\begin{equation}
\label{eq:R_1_def}
R_1 \simeq \frac{10^{-8/3}}{\Lambda}\left(\frac{M_{\rm pl}}{\Lambda}\right)^{2/3}.
\end{equation}
The initial radius at the beginning of the contraction stage is:
\begin{equation}
\label{eq:Ri_contraction}
R_i =\textrm{Max}\left[R_0,R_1 \right],
\end{equation}
where $R_0$ is taken from the numerical studies reported in Eq.~\eqref{eq:R_0_def}.
One finds that the bubble radius is set by coalescence, $R_i\simeq R_1$, only for $\Lambda \gtrsim \rm TeV$.

\subsection{Pocket contraction}
\label{app:pocket_contraction}
Replacing the bubble radius by the pocket radius in Eq.~\eqref{eq:dvdR}, $dv_w/dR$ becomes $-dv_w/dR$ and one obtains:
\begin{equation}
\label{eq:dvdR_pocket}
    \frac{dv_w}{dR}=-\frac{H}{v_w}+\frac{3 L R^2}{8\rho R_0^3}.
\end{equation}
Around the time of percolation, the constant temperature solution $\dot{T}\simeq 0$ in Eq.~\eqref{eq:vw_instant_therm} which corresponds to the solution ${dv_w}/{dR}\propto\dot{T}\simeq 0$ in Eq.~\eqref{eq:dvdR_pocket} still applies with now $R$ being the pocket radius instead of the bubble radius.\footnote{Eq.~\eqref{eq:dvdR_pocket} can also be derived from Eq.~\eqref{eq:T_dot} with  $x=1-\frac{R^3}{2R_0^3}$.}
The second term in Eq.~\eqref{eq:dvdR_pocket} goes to zero as the pocket shrinks $R\to 0$. Below some radius $R_f$, the left hand side $dv_w/dR$ can not be neglected anymore, and Eq.~\eqref{eq:dvdR_pocket} is better approximated by
\begin{equation}
\frac{dv_w}{dR} \simeq -\frac{H}{v_w},
\end{equation}
which after integration gives the value of the bubble wall velocity during the rest of the contraction period
\begin{equation}\label{eq:vwR_2}
    v_w\simeq c_0 \sqrt{2H R_f},
\end{equation}
where $c_0\simeq 2$ is a correcting factor whose value is fitted on numerical results from \cite{Asadi:2021pwo}.
The value of $R_f$ can be obtained by matching Eqs.~(\ref{eq:vwR_2}) with (\ref{eq:vw_instant_therm}),
\begin{equation}
    R_f \simeq 2 H^{1/5} R_i^{6/5}.
\end{equation}
Plugging back into Eq.~\eqref{eq:vwR_2} leads to
\begin{equation}
\label{eq:pocket_velocity_app}
v_w \simeq \begin{cases}
2\times 10^{-3}\left( \frac{\Lambda}{M_{\rm pl}}\right)^{0.06},\qquad ~R_i =R_0,\\
0.2\left( \frac{\Lambda}{M_{\rm pl}}\right)^{1/5},\qquad \qquad \quad R_i =R_1,
\end{cases}
\end{equation}
where $R_0$ and $R_1$ are given in Eqs.~\eqref{eq:R_0_def} and \eqref{eq:R_1_def}.
Only the second line was used in \cite{Asadi:2021pwo}.

\section{Dark baryon abundance}
\label{sec:dark_baryon_abundance}
\subsection{Quark freeze-out}
\label{sec:quark_FO}

The evolution of the quark abundance before the onset of the confining phase transition is described by the Boltzmann equation:
\[ \frac{dY_{\mathcal{Q}}}{dx} = -\frac{\lambda_n}{x^{2+n}} \left( Y_{\mathcal{Q}}^2 -  Y_{\mathcal{Q}, eq}^2 \right), \qquad \lambda_n \equiv M_{\rm pl} m_{\mathcal{Q}} \, \sigma_n \, \sqrt{8\pi^2 \gSM /45},\]
where $x\equiv m_{\mathcal{Q}}/T$, $\sigma_n \equiv \langle \sigma_{\rm ann} v_{\mathsmaller{\rm rel}}\rangle\, x^{n}$ with $n$ chosen so that $\sigma_{n}$ is $x$ independent, and $M_{\rm pl} \simeq 2.44 \times 10^{18} ~ \text{GeV}$. For s-wave, one has $n=0$ or $n=-1/2$ with or without Sommerfeld effects
 As the temperature goes below \(m_{\mathcal{Q}}\), DM predominantly annihilate into lighter species. This continues until the annihilation rate falls below the universe expansion rate, leading to a frozen-out DM abundance around temperature, \(T_{\rm FO}\), defined by \cite{Kolb:1990vq}:
\[ \xFO = \log[0.192(n+1) M_{pl} m_{\mathcal{Q}} \sigma_n  \gDM/\gSM] - (n+0.5)\log[\xFO]. \]
where \( \gDM \) represents DM degrees of freedom.  The present abundance is derived from the previously frozen abundance redshifted up to today:
\[ \Omega_{\mathcal{Q}} h^2 = 2\frac{s_0 m_{\mathcal{Q}}}{3 M_{pl}^2 H_{100}^2}  Y_{\mathcal{Q}}^{\rm FO}. \]
where $H_{100} = 100$~km/s/Mpc, $s_0=2891.2$~cm$^{-3}$~\cite{ParticleDataGroup:2022pth} and:
\begin{equation}
\label{eq:YDM_FO}
     Y_{\mathcal{Q}}^{\rm FO} = \frac{(n+1) x_{\rm FO}^{n+1}}{\lambda_n}. 
\end{equation}
The factor $2$ in the abundance equation considers quarks population in the sum of its two parts: \(\Omega_{\mathcal{Q}}\) and \(\Omega_{\overline{\mathcal{Q}}}\).
The quark annihilation cross-section reads:
\begin{equation}
\label{eq:sigmav_ann_app}
\left<\sigma_{\rm ann} v_{\rm rel}\right> = \zeta\frac{\pi \alpha_{\rm DC}^2}{m_{\mathcal{Q}}^2},
\end{equation}
where $\zeta$ accounts for non-perturbative effects \cite{ArkaniHamed:2008qn} which become large when $\alpha_{\rm DC}/v_{\rm rel} \gg 1$. Assuming quark and anti-quark in fundamental representation of $SU(N_{\rm DC})$, non-perturbative effects operate through two channels, an attractive singlet and a repulsive adjoint \cite{Mitridate:2017oky}:
\begin{equation}
\zeta = \frac{1}{N_f}\frac{(N_{\rm DC}^4-3N_{\rm DC}^2+2)}{16N_{\rm DC}^3}\left(\frac{2}{N_{\rm DC}^2-2} S_1 + \frac{N_{\rm DC}^2-4}{N_{\rm DC}^2-2}S_{\rm adj} \right),
\end{equation}
where $S_J$ is the Sommerfeld factor \cite{Sommerfeld:1931qaf}:
\begin{equation}
S_J= \frac{2\pi \alpha_{\rm eff,J}/v_{\rm rel}}{1-e^{-2\pi \alpha_{\rm eff,J}/v_{\rm rel}}},
\end{equation}
and $\alpha_{\rm eff,J} = \lambda_J \alpha_{\rm DC}$ with $\lambda_J=C_{R_J}+C_{R_J}-C_{J}$ where $C_J$ are quadratic Casimir of the initial particles and the resulting bound-state, $C_J = (N_{\rm DC}^2-1)/2N_{\rm DC}$ and $N_{\rm DC}$ for fundamental and adjoint representation of $SU(N_{\rm DC})$ respectively. For $N_{\rm DC} = 3$, we have
\begin{equation}
\zeta = \frac{7}{54N_f}\left(\frac{2}{7} S_1 + \frac{5}{7}S_8 \right),\qquad \lambda_1=4/3, \qquad \lambda_8 = -1/6,
\end{equation}
where $N_f=1$ if dark quarks are SM singlet and $N_f=2$ or $3$ if they are $SU(2)_L$ doublet or triplet.
 We neglect additional channels due to dark quarks potentially having SM charges. During freeze-out $v_{\rm rel}\simeq 2\sqrt{2T_{\rm FO}/\pi m_{\mathcal{Q}}} \sim 0.3$ and $\alpha_{\rm DC} \sim 0.17$, where we plugged $T_{\rm FO}\sim m_{\mathcal{Q}}/30$ in Eq.~\eqref{eq:alpha_D_rge}. Hence $\alpha_{\rm DC}/v_{\rm rel} \sim 0.6 \lesssim 1$ and non-perturbative effects are small and we can safely neglect the contribution of bound-state formation to the freeze-out abundance \cite{vonHarling:2014kha}. Instead during the completion of the $SU(N_{\rm DC})$ phase transition, taking for example $m_{\mathcal{Q}} \sim 10^4\Lambda$,  the quark velocity is much smaller  $v_{\rm rel}\sim 0.016$ and $\alpha_{\rm DC} \sim 0.06$ such that $\alpha_{\rm DC}/v_{\rm rel} \sim 3.9 \gtrsim 1$. Hence, non-perturbative effects due to the long range of the dark force can be active during the phase transition. Effects from bound-state formation during the phase transition are discussed in the next section.

\subsection{Quark squeeze-out}
\label{app:Quark_SO}

\begin{figure*}[ht!]
    \centering
    \includegraphics[width=0.7\textwidth]{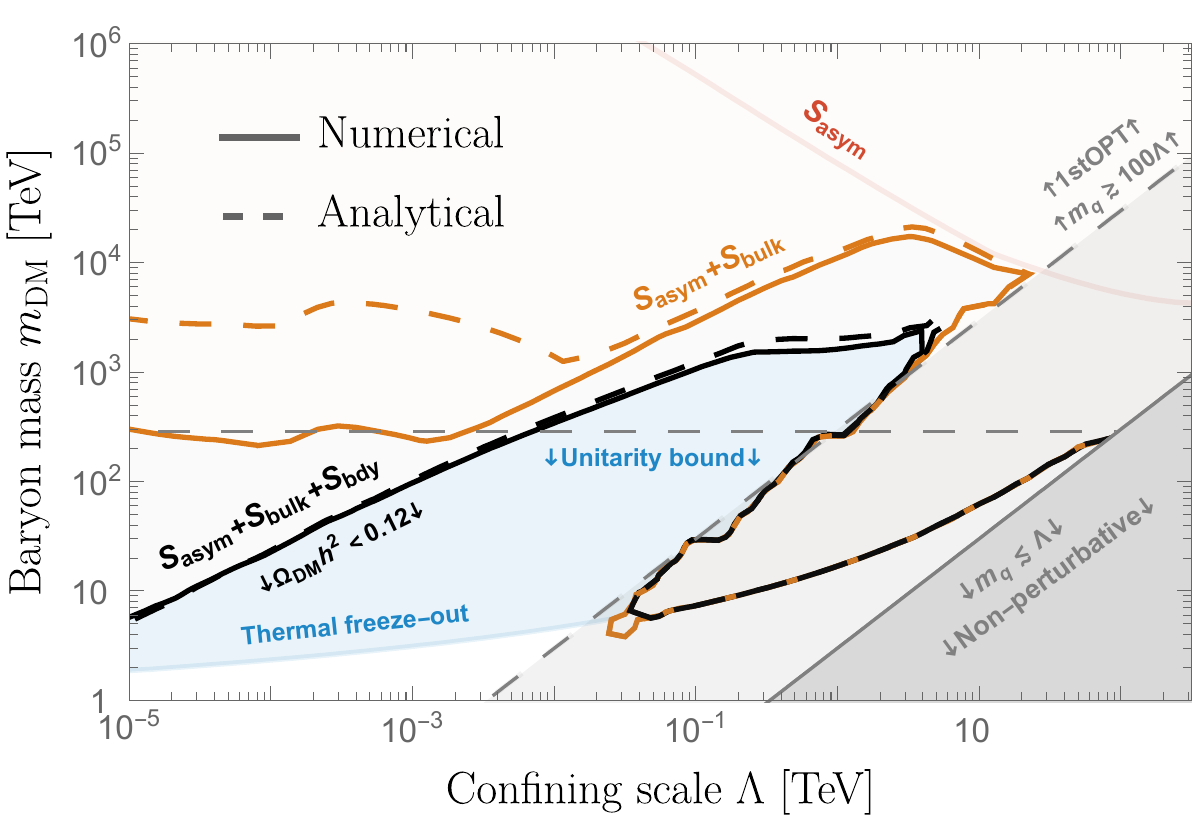}
    \caption{We evaluate the DM relic abundance using an analytical formula (highlighted as \textbf{Dashed}) and compare it with the results obtained from the numerical integration of the Boltzmann equations set (denoted as \textbf{Solid}). The agreement is rather good, both in the context of baryon formation through bulk contribution, given by the first piece in  Eq.~\eqref{eq:surv_S_symm} (indicated in \textbf{Orange}) and through combined bulk and boundary contributions given by the two pieces in Eq.~\eqref{eq:surv_S_symm} (represented in \textbf{Black}), where $N_3$ is obtained from algebraically solving Eqs.~\eqref{eq:quasi_eq_1} and \eqref{eq:quasi_eq_2}.}
    \label{fig:ana_vs_num}
\end{figure*}

\textbf{Quark survival factor.}
During the phase transition, single heavy quark entering the bubble form flux tube attached to the wall. Nucleation of quark anti-quark pairs is exponentially suppressed for $m_{\mathcal{Q}} \gg \Lambda$. Hence the flux-tube is stable and the quark are sling-shot back to the deconfined phase. This leads to an accumulation of quarks inside pockets outside bubbles.
The shrinking of pockets switches on a second stage of quark annihilation. Quarks can survive this dynamics through three ways: from the accidental asymmetry in each pocket, by forming baryons in the bulk which subsequently exit the pocket, by forming baryons directly at the boundary. We express those three contributions in terms of the quark survival factor:
\begin{equation}
\label{eq:survival_fac_app}
    S \equiv \frac{3N_{3}^f}{N_1^i}=  S_{\rm asym}+ S_{\rm bulk}+ S_{\rm bdy},
\end{equation}
where the initial number of quarks per pockets reads
\begin{equation}
\label{eq:quark_abundance_initial}
    N_1^i =  2 Y_\mathcal{Q}^{\rm FO} s \frac{4\pi}{3}R_i^3,
\end{equation}
where $Y_\mathcal{Q}^{\rm FO}$ in Eq.~\eqref{eq:YDM_FO}, $R_i$ is the pocket radius at the beginning of the contraction stage, $ s=2\pi^2g_{*}T_c^3/45$ is the universe entropy density, and the factor $2$ assumes that pockets occupy half the total universe volume.\\
\begin{figure*}[ht!]
    \centering \includegraphics[width=0.7\textwidth]{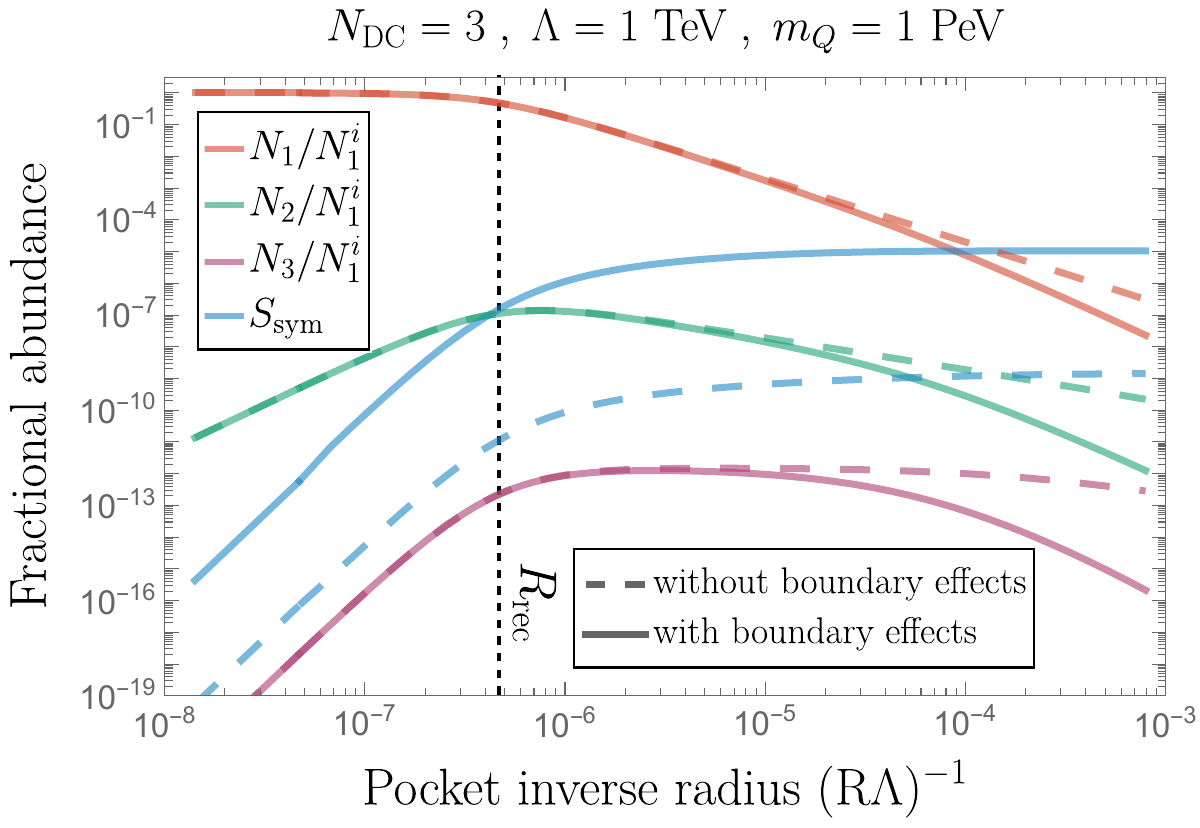}
    \caption{Evolution of the number of quarks $N_1$ (\textbf{Red}), di-quarks $N_2$ (\textbf{Green}) and baryons $N_3$ (\textbf{Purple}) inside the pocket during its contraction stage resulting from numerically integrating the set of Boltzmann equations in Eqs.~\eqref{eq:boltz_full_N1}, \eqref{eq:boltz_full_N2} and \eqref{eq:boltz_full_N3}. Above the recoupling radius $R\gtrsim R_{\rm rec}$, the quark number $N_1$ is approximately constant while di-quark and baryon are constantly produced. Below the recoupling radius $R\lesssim R_{\rm rec}$, quarks annihilate efficiently with antiquarks and they number decrease. As a consequence, the production of di-quark and baryons stops. The symmetric component of the survival abundance $S$ of quarks under the form of baryons which have escaped the pocket (\textbf{Blue}), defined in Eq.~\eqref{eq:S_symm_def_app}, increases at a constant rate above $R \gtrsim R_{\rm rec}$ and freezes-out below $R \lesssim R_{\rm rec}$. In presence of baryon formation directly at the boundary (\textbf{Solid} vs \textbf{Dashed}), the quark surviving abundance $S$ is much larger. Also in presence of boundary effects, the number of species $N_1$, $N_2$ and $N_3$ decrease faster at the end of the contraction stage.}
    \label{fig:fractional_abundance_app}
\end{figure*}
\textbf{Asymmetric fraction.}
  Let's introduce the total number $N$ of quarks and anti-quark and their difference $\Delta$ inside a given pocket:
 \beq
N  \equiv \left<N_{\mathcal{Q}} +N_{\bar{\mathcal{Q}}}\right>, \qquad \Delta \equiv N_{\mathcal{Q}} -N_{\bar{\mathcal{Q}}},
\eeq
where $N_{\mathcal{Q}} \equiv N_1^i$ and $N_{\bar{\mathcal{Q}}} \equiv N_{-1}^i$ refer to the initial numbers of quarks and antiquarks.
The entropy of the quarks and antiquarks in the pocket reads:
\beq
S(\Delta; N) =
\log\left(\frac{N!}{N_{\mathcal{Q}}!(N-N_{{\mathcal{Q}}})!}\right) \simeq N \log 2 -\frac{\Delta^2}{2N}\,,
\eeq
where we used Stirling's approximation $\log{n!}\simeq n\log{n}-n$ and Taylor-expanded in the limit $\Delta \ll N$. In the absence of substantial interactions between quark and antiquarks, the energy of each microstates is independent of $\Delta$.  Hence, the probability distribution of the quark/antiquark asymmetry $\Delta$ reads:
\begin{equation}
    P({\Delta};N) = \frac{e^{S({\Delta};N)}}{Z(N)},
\end{equation}
where $Z(N)= \sum_{N_Q=0}^{N}\frac{N!}{N_{\mathcal{Q}}!(N-N_{{\mathcal{Q}}})!}=2^N$ is the partition function. 
$P({\Delta};N)$ is a Gaussian with mean $\left<\Delta\right>=0$ and standard deviation:
\begin{equation}
    \sqrt{\left<\Delta^2\right>}\equiv \sqrt{\frac{\sum_{\Delta} \Delta^2 P({\Delta};N) }{\sum_{\Delta}P({\Delta};N)}}=\sqrt{N}.
\end{equation}
We deduce that the asymmetric contribution to the quark survival factor $S\equiv N_1^f/N_1^i$ in Eq.~\eqref{eq:survival_fac_app} is given by:
\begin{equation}
  S_{\rm asym} = \frac{1}{2}\frac{ \sqrt{\left<\Delta^2\right>}}{N_{\mathcal{Q}}} = \frac{1}{2}\frac{\sqrt{2N_1^i}}{N_1^i} = \frac{1}{\sqrt{2N_1^i}}.
\end{equation}
The factor $1/2$ in the first equality is `added by-hand' to compensate for the factor of $2$ accounting for the symmetric component of antibaryons in the DM abundance formula in Eq.~\eqref{eq:Y_DM_after_PT}. The factor of $\sqrt{2}$ in the numerator of the second equality arises from $\sqrt{N}=\sqrt{N_{\mathcal{Q}}+N_{\bar{\mathcal{Q}}}}\simeq \sqrt{2N_{\mathcal{Q}}}$. Both factors were omitted in \cite{Asadi:2021pwo,Asadi:2021yml}. \\
\textbf{Symmetric fraction.}
The symmetric component $S_{\rm sym}=S_{\rm bulk}+ S_{\rm bdy}$ of the quark survival fraction can be found from evaluating the two integrals:
\begin{equation}
\label{eq:S_symm_def_app}
S_{\rm bulk} = \frac{1}{v_w N_1^i}\int_{\rm R_i}^0 dR\, \Gamma_{3 \rightarrow 3}^{\rm esc}, \quad \textrm{and} \quad S_{\rm bdy} = \frac{1}{v_w N_1^i}\int_{\rm R_i}^0 dR\left(3\Gamma_{(1,1,1) \rightarrow 3}^{\rm bdy}+\Gamma_{(1,2) \rightarrow 3}^{\rm bdy}\right).
\end{equation}
The evolution of the number of quarks $N_1$, di-quarks $N_2$ and baryons $N_3$ is set by the following set of Boltzmann equations:
\begin{align}\label{eq:boltz_full_N1}
\dot{N}_1 =&-\Gamma^{\rm bulk}_{(-3,1)\to(-1,-1)}-\Gamma^{\rm bulk}_{(-3,1)\to(-2,0)}+2\Gamma^{\rm bulk}_{(3,-1)\to(1,1)}+\Gamma^{\rm bulk}_{(3,-2)\to(1,0)}-\Gamma^{\rm bulk}_{(1,-1)\to(0,0)} \nonumber\\ &
+\Gamma^{\rm bulk}_{(2,2)\to(3,1)}-2 \Gamma^{\rm bulk}_{(1,1)\to(2,0)} +\Gamma^{\rm bulk}_{(-3, 2)\to(-2, 1)}+ \Gamma^{\rm bulk}_{(2, -2)\to(1,-1)}-\Gamma^{\rm bulk}_{(2,1)\to(3,0)} + \Gamma^{\rm bulk}_{(2,-1)\to(1,0)} \nonumber \\ 
&-\Gamma^{\rm bulk}_{(-2,1)\to(-1,0)}+\Gamma^{\rm bulk}_{(3,-3)\to(-1,1)}  -\Gamma_{(1,-1) \rightarrow \mathcal{M} }^{\rm bdy} - 3\Gamma_{(1,1,1) \rightarrow 3}^{\rm bdy} -\Gamma_{(1,2) \rightarrow 3}^{\rm bdy}\, ,   \\
\label{eq:boltz_full_N2}
\dot{N}_2=& - \Gamma^{\rm bulk}_{(-3,2)\to(-1,0)}+ \Gamma^{\rm bulk}_{(3,-1)\to(2,0)}- \Gamma^{\rm bulk}_{(2, -2)\to(0,0)}+\Gamma^{\rm bulk}_{(3,-2)\to(2,-1)}+\Gamma^{\rm bulk}_{(3,-3)\to(2,-2)}-2\Gamma^{\rm bulk}_{(2, 2)\to(3,1)}\nonumber\\
& - \Gamma^{\rm bulk}_{(2,1)\to(3,0)} - \Gamma^{\rm bulk}_{(2,-1)\to(1,0)} + \Gamma^{\rm bulk}_{(1,1)\to(2,0)}-\Gamma^{\rm bulk}_{(-3,2)\to(-2,1)}- \Gamma^{\rm bulk}_{(2,-2)\to(1,-1)}-\Gamma_{(1,2) \rightarrow 3}^{\rm bdy},\\ 
\label{eq:boltz_full_N3}
\dot{N}_3 =&\; \Gamma^{\rm bulk}_{(2, 2)\to(3,1)}-\Gamma^{\rm bulk}_{(3, -3)\to(0, 0)} - \Gamma^{\rm bulk}_{(3,-1)\to(2,0)} + \Gamma^{\rm bulk}_{(2,1)\to(3,0)}-\Gamma^{\rm bulk}_{(3, -1)\to(1,1)}-\Gamma^{\rm bulk}_{(3, -3)\to(1, -1)} \nonumber\\
&-\Gamma^{\rm bulk}_{(3, -3)\to(2, -2)}-\Gamma^{\rm bulk}_{(3, -2)\to(2, -1)}-\Gamma^{\rm bulk}_{(3, -2)\to(1, 0)}-\Gamma_{3 \rightarrow 3}^{\rm esc}\,.
\end{align}
The time derivative can be traded by a radius derivative through $\dot{N}_i=-v_wN_i'$.
From top to bottom, left to right, the boundary terms  $\Gamma^{\rm bdy}_{\rm A \to B}$ include quark / anti-quark annihilation, baryon formation from three quarks, from quark / di-quark, and baryon escape rate (noted `esc'), all occurring at the wall boundary. As they are ones of the novelties of this work, we discuss them in details in the next paragraph.
Instead, the bulk interaction rates read:
\begin{equation}
    \label{eq:finiteblz_app}
   \Gamma^{\rm bulk}_{a+b \to c+d}= \sum_{\rm a+b = c+d}s_{\rm ab,cd}^i \frac{\langle\sigma v\rangle_{abcd}^{\rm bulk}}{V}\left(N_aN_b-N_cN_d f_{abcd}\right),
\end{equation}
where $s_{\rm ab,cd}^i$ is the net number of $i$ particles destroyed in the process $ab \to cd$ with cross-section $\left<\sigma v\right>_{\rm ab cd}$ and:
\begin{equation}
\label{eq:f_abcd}
    f_{abcd}\equiv n^{\rm eq}_a n^{\rm eq}_b/(n^{\rm eq}_c n^{\rm eq}_d),
\end{equation}
 with $n_x^{\rm eq}$ the number densities at equilibrium.
The different bound-state cross-sections $\langle\sigma v\rangle_{abcd}^{\rm bulk}$ for the bulk processes are extracted from the notebooks used in \cite{Asadi:2021pwo}. 
Anticipating the derivation of boundary terms in the next paragraph, we can already have a look in Fig.~\ref{fig:fractional_abundance_app} at the results from numerically integrating the set of Boltzmann equations in Eqs.~\eqref{eq:boltz_full_N1}, \eqref{eq:boltz_full_N2} and \eqref{eq:boltz_full_N3}. We can see that the specie numbers in the pocket follow the hierarchy $N_1\gg N_2 \gg N_3$.
For the purpose of deriving analytical results later on, we can use this hierarchy to get rid of sub-dominant bulk interactions in the set of Boltzmann equations in Eqs.~\eqref{eq:boltz_full_N1}, \eqref{eq:boltz_full_N2} and \eqref{eq:boltz_full_N3}. One obtains:
\begin{align}\label{eq:boltz_app}
\dot{N}_1 &\simeq -\Gamma^{\rm bulk}_{(1,-1)\to(0,0)} -2 \Gamma^{\rm bulk}_{(1,1)\to(2,0)}   -\Gamma^{\rm bulk}_{(2,1)\to(3,0)}-\Gamma^{\rm bulk}_{(-2,1)\to(-1,0)} + \Gamma^{\rm bulk}_{(2,-1)\to(1,0)}   \nonumber\\
&\;\;\;\;-\Gamma_{(1,-1) \rightarrow \mathcal{M} }^{\rm bdy}- 3\Gamma_{(1,1,1) \rightarrow 3}^{\rm bdy} -\Gamma_{(1,2) \rightarrow 3}^{\rm bdy}\, ,   \\
\label{eq:boltz_app_N2}
\dot{N}_2&\simeq  - \Gamma^{\rm bulk}_{(2,1)\to(3,0)} - \Gamma^{\rm bulk}_{(2,-1)\to(1,0)} + \Gamma^{\rm bulk}_{(1,1)\to(2,0)}  -\Gamma_{(1,2) \rightarrow 3}^{\rm bdy},\\ 
\label{eq:boltz_app_N3}
\dot{N}_3 &\simeq  - \Gamma^{\rm bulk}_{(3,-1)\to(2,0)} + \Gamma^{\rm bulk}_{(2,1)\to(3,0)} -\Gamma_{3 \rightarrow 3}^{\rm esc}\,.
\end{align}
We now proceed to the derivation of boundary terms. \\
\begin{figure}[!ht]
\centering
\raisebox{0cm}{\makebox{\includegraphics[width=0.8\textwidth, scale=1]{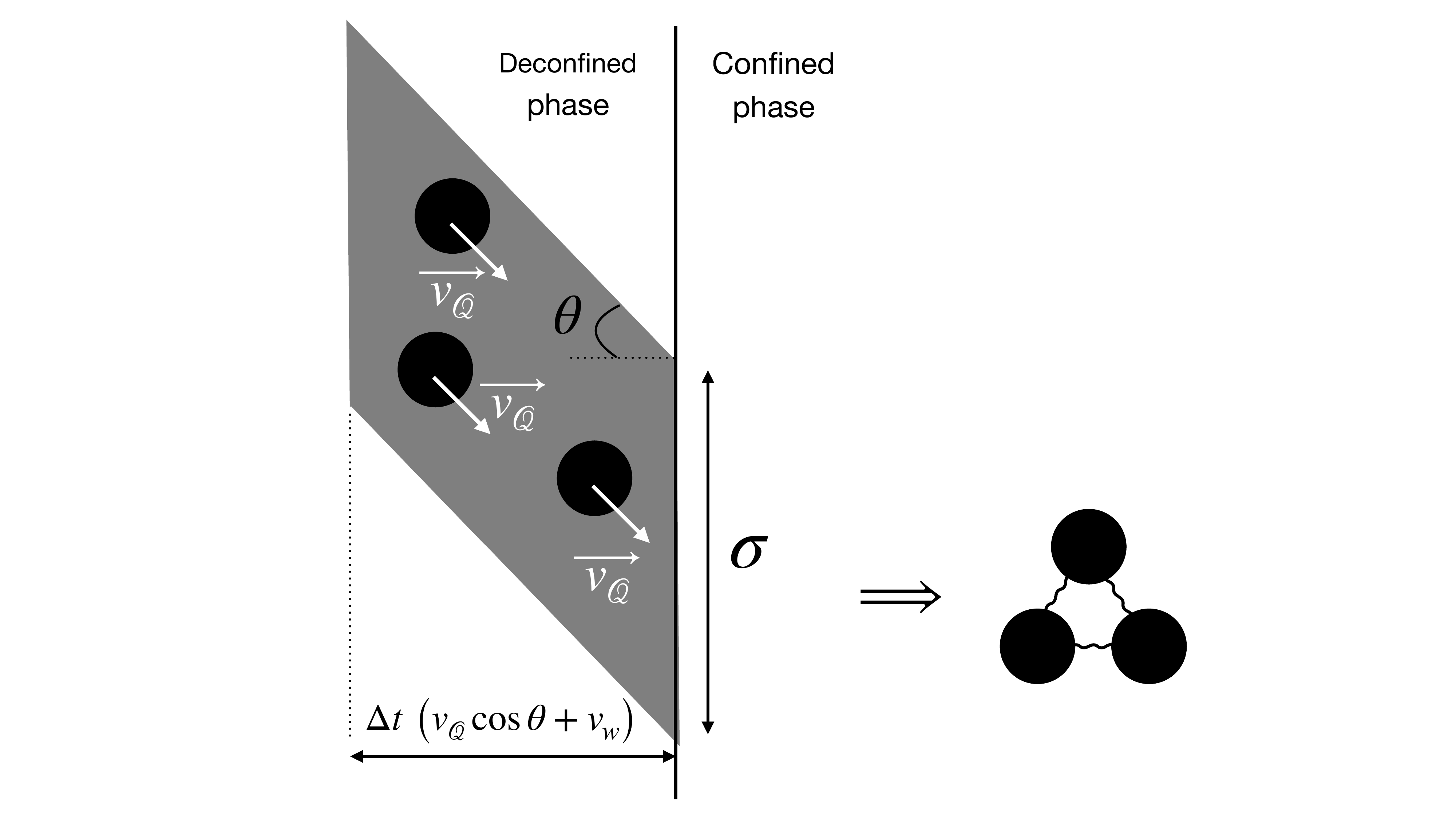}}}
\caption{\it \small In order to form a bound-states at the wall boundary, a color-neutral configuration (here a baryon with $N_{\rm DC}=3$ colors) must pass through the same surface area $\sigma \simeq \pi \ell^2$ during $\Delta t \simeq \ell/v_{\mathcal{Q}}$ with $\ell \simeq 1/\Lambda$.}
\label{fig:drawing_bdy} 
\end{figure}
\FloatBarrier
\textbf{Boundary Effects.}
When a colored particles in the bulk encounter the wall boundary, it will form a flux tube attached to the wall. As the particle penetrate inside the confined phase, the flux tube will act a force pulling the particle back to the deconfined phase.
In order to escape the pocket through the boundary, a color-singlet assembly of charged particles must collectively impinge the same surface area $\sigma$ during the same time $\Delta t$, see Fig.~\ref{fig:drawing_bdy}.
We take:
\begin{equation}
\label{eq:BSF_cross-section_wall}
    \sigma =  f_\sigma \pi\ell^2,\qquad \ell \simeq 1/\Lambda,
\end{equation}
where $\ell$ is the typical flux tube width and $f_\sigma$ a form factor encoding uncertainties in the modelling. We assume $f_\sigma$ to be identical for all boundary processes. The quantity $\Delta t$ is the typical time for charged species to remain attached to the wall by a flux tube acting with a force $F\simeq \Lambda^2$ before bouncing off \cite{Asadi:2021pwo}:
\begin{equation}
    \Delta t \simeq \frac{\bar{v}}{\dot{v}_q} \simeq  \frac{v_{\mathcal{Q}}+v_w}{\Lambda^2/m_{\mathcal{Q}}} \simeq \frac{v_{\mathcal{Q}}+v_w}{v_{\mathcal{Q}}^2}\ell ~\overset{v_{\mathcal{Q}}\gg v_w}{\simeq} ~\frac{\ell}{v_{\mathcal{Q}}}, 
\end{equation}
where  $\bar{v} = v_{\mathcal{Q}} + v_w$ is the average particle velocity in the wall frame with $v_{\mathcal{Q}} \simeq \sqrt{\Lambda/m_{\mathcal{Q}}} \gg v_w$ and $\ell\simeq 1/\Lambda$. Equivalently, $\Delta t \simeq d/(v_{\mathcal{Q}}+v_w)$ where $d\simeq E_{\rm kin}/F$ is the distance from the wall at which outgoing particles with initial kinetic energy $E_{\rm kin}\simeq \Lambda (v_{\mathcal{Q}}+v_w)^2/v_{\mathcal{Q}}^2$ in the wall frame turn back under the effect of the confining force $F\simeq \Lambda^2$. 
 The flux of particles passing through the wall is $n_i \bar{v}_\perp$ where $\bar{v}_\perp$ is the average of the normal component of the particle velocity in the wall frame:
\begin{equation}
 \bar{v}_\perp =v_w+\frac{1}{4\pi}\int_0^{2\pi}d\phi \int_0^{\pi/2}  v_{\mathcal{Q}}\cos(\theta)\sin\theta d\theta =v_w + \frac{v_{\mathcal{Q}}}{4} \simeq \frac{v_{\mathcal{Q}}}{4}
\end{equation}
We deduce the mean number of particles passing through $\sigma$ during $\Delta t$:
\begin{equation}
\label{eq:lambda_i_def}
\lambda_i = \Delta t \,\bar{v}_\perp\, n_i \,\sigma \simeq f_\sigma\frac{\pi}{4} n_i \ell^3 \simeq f_\sigma \frac{3N_i}{16(R\Lambda)^3},
\end{equation}
where we used $n_i=N_i/V$ and $V=4\pi R^3/3$.
Assuming that the particles do not influence each others arrivals on the wall, the number of particles impinging the wall in the same window of time $\Delta t$ is given by a Poisson distribution. The probability that $k$ particles of type $i$ meet on the wall is:
\begin{equation}
\label{eq:Poisson_num}
   P_{\lambda_i}\left(k\right)=\frac{\lambda_i^k}{k!}e^{-\lambda_i}\;.
   \end{equation}
 We successfully tested the Poisson distribution in Eq.~\eqref{eq:Poisson_num} against a numerical simulation of $N_i= 10^8$ quarks randomly distributed in a $L_x\times L_y\times L_z$ box. We gave the quark velocity a fixed norm $v_{\mathcal{Q}}$ and a random direction. After a time interval $\Delta t$, some quarks cross over the wall in the $x-y$ plan. We fully cover the wall by $\ell\times\ell$ targets and count the number of the quarks meeting in the same target. We were able to successfully recover the Poisson distribution.
We are now able to calculate the different boundary terms appearing in the Boltzmann equations.
The rate at which 3 quarks in the bulk form a baryon at the boundary is: 
\begin{align}
\label{eq:gammaNc}
\Gamma_{(1,1,1) \rightarrow 3}^{\rm bdy}&=\oint_{\partial V}dS\frac{1}{\sigma\Delta t}\sum_{k=N_{\rm DC}}^\infty P_{\lambda_1}\left(k\right) \notag \\
&=\frac{4\pi R^2}{\sigma\Delta t} \left[1-\frac{\Gamma(N_{\rm DC},\lambda_1)}{\Gamma(N_{\rm DC})}\right] \notag \\
&\simeq \frac{4 v_{\mathcal{Q}} \lambda_1^{N_{\rm DC}}R^2}{f_\sigma\ell^3\Gamma(N_{\rm DC}+1)}\; \notag \\
&\simeq v_{\mathcal{Q}}\frac{9 f_\sigma^2 }{2048 \Lambda^6} \frac{N_1^3}{R^7} .
\end{align}
Eq.~\eqref{eq:gammaNc} is the Taylor-expanded form in the limit $\lambda_1\ll 1$. Instead for $\lambda_1 \gg 1$, the Poisson sum $\sum_{k=N_{\rm DC}}^\infty P_{\lambda_1}\left(k\right) $ converges to $1$ and we get the upper bound $\Gamma_{(1,1,1) \rightarrow 3}^{\rm bdy}\leq 4 v_{\mathcal{Q}} R^2\Lambda^3$.
We also compute the rate for quark/anti-quark pair forming a meson:
\begin{align}
    \Gamma_{(1,-1) \rightarrow \mathcal{M}}^{\rm bdy}&=\oint_{\partial V}dS\frac{1}{\sigma\Delta t} \left[\sum_{k=1}^\infty P_{\lambda_1}\left(k\right) \right]^2\notag \\
&=\frac{4\pi R^2}{\sigma\Delta t} \left[1-\Gamma(1,\lambda_1)\right]^2 \notag \\
&\simeq  \frac{4v_{\mathcal{Q}}\lambda_1^2R^2}{f_\sigma\ell^3} \notag\\
\label{eq:1m1-meson}
&\simeq v_{\mathcal{Q}}\frac{9f_\sigma}{64 \Lambda^3}\frac{N_1^2}{R^4},
\end{align}
the rate of quark / di-quark pair forming a baryon:
\begin{align}
    \Gamma_{(1,2) \rightarrow 3}^{\rm bdy}&=\oint_{\partial V}dS\frac{1}{\sigma\Delta t} \left[\sum_{k=1}^\infty P_{\lambda_1}\left(k\right) \right]\left[\sum_{k=1}^\infty P_{\lambda_2}\left(k\right) \right]\notag \\
&=\frac{4\pi R^2}{\sigma\Delta t} \left[1-\Gamma(1,\lambda_1)\right] \left[1-\Gamma(1,\lambda_2)\right] \notag \\
&\simeq \frac{4v_{\mathcal{Q}}\lambda_1\lambda_2R^2}{f_\sigma\ell^3}\;\quad\notag\\
&\simeq v_{\mathcal{Q}}\frac{9f_\sigma}{64 \Lambda^3}\frac{N_1N_2}{R^4},
\end{align}
the rate for the quarks of baryons in the bulk simply passing through the wall:
\begin{align}
    \Gamma_{3 \rightarrow 3}^{\rm esc} &= \oint_{\partial V} \hspace{-0.2cm}dS \frac{1}{\sigma \Delta t} \left[\sum_{k=1}^\infty k P_{\lambda_3}\left(k\right) \right]\notag\\
    &= \frac{4 v_{\mathcal{Q}}\lambda_3R^2}{f_\sigma\ell^3} \notag \\
    &= \frac{3v_{\mathcal{Q}} N_3}{4R}. 
\end{align}
\begin{figure*}[t!]
    \centering
    \includegraphics[width=0.48\textwidth]{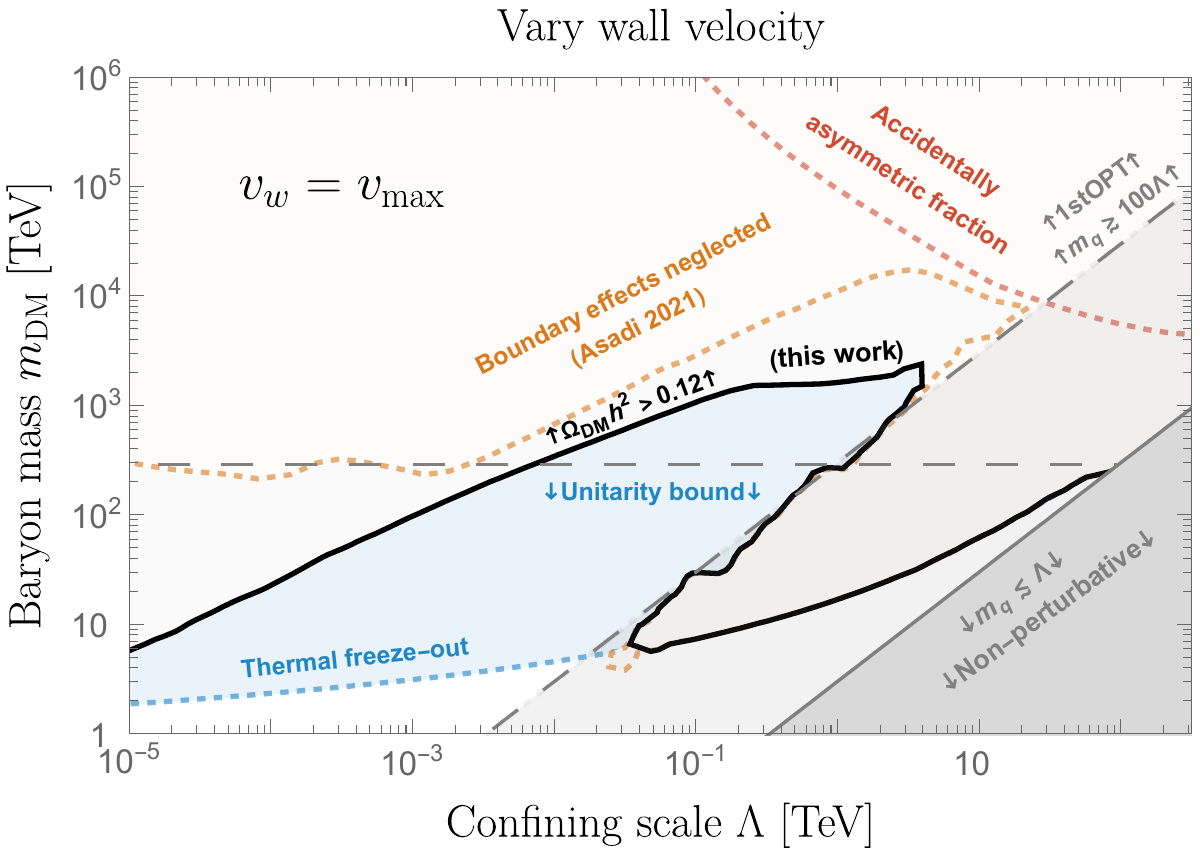}
    \includegraphics[width=0.48\textwidth]{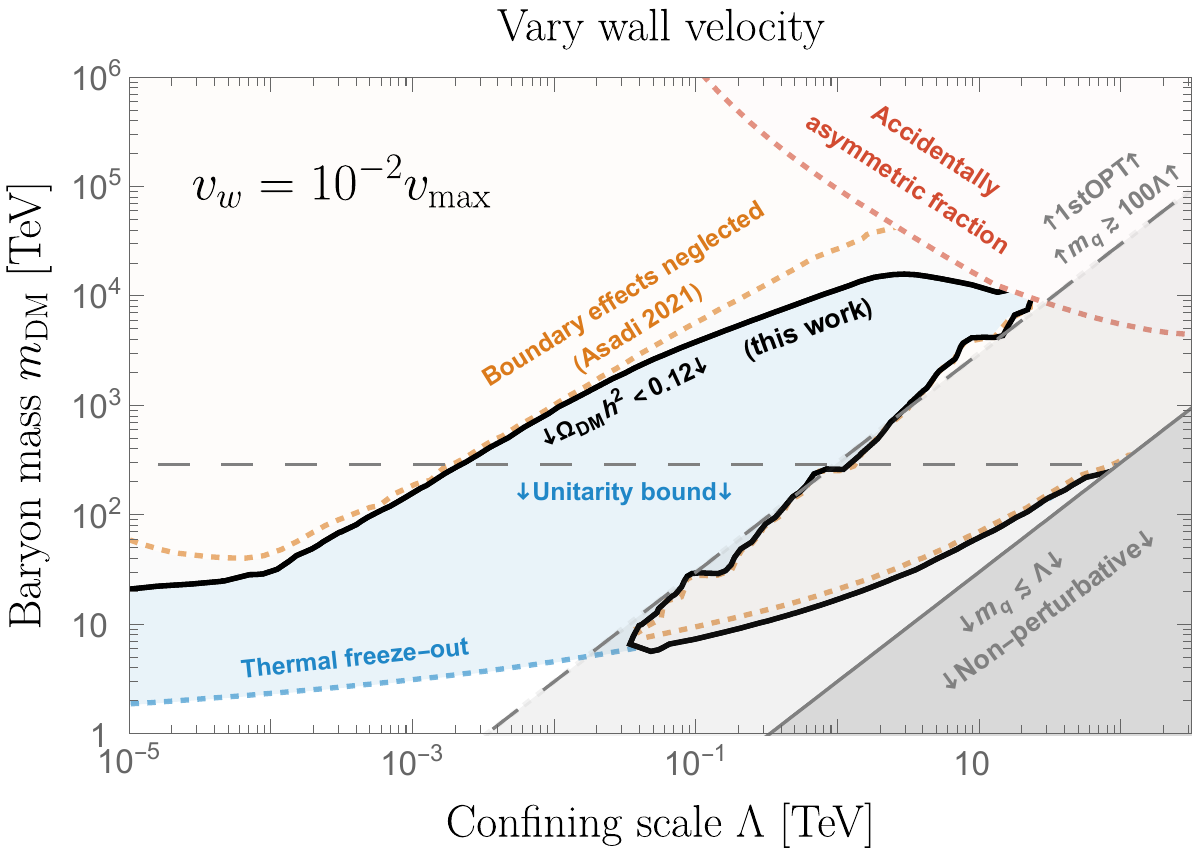}
    \includegraphics[width=0.48\textwidth]{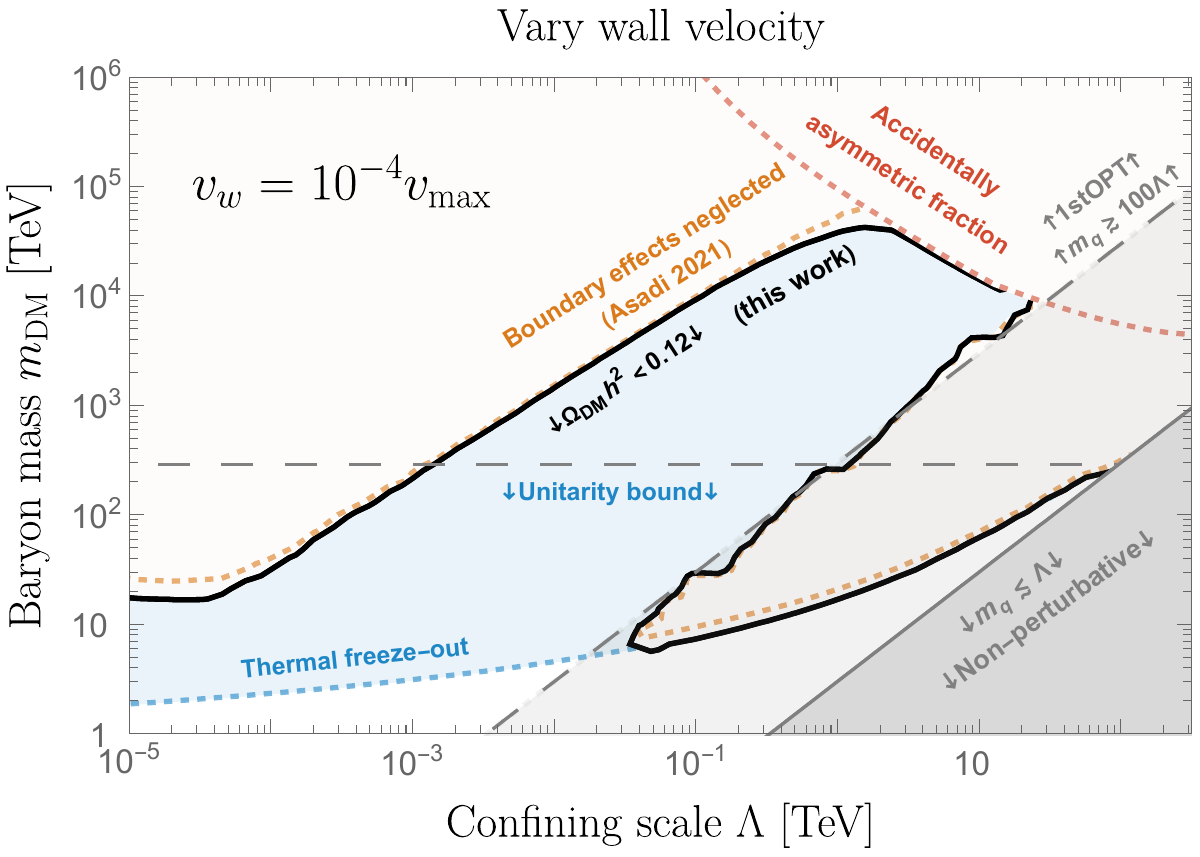}
    \includegraphics[width=0.48\textwidth]{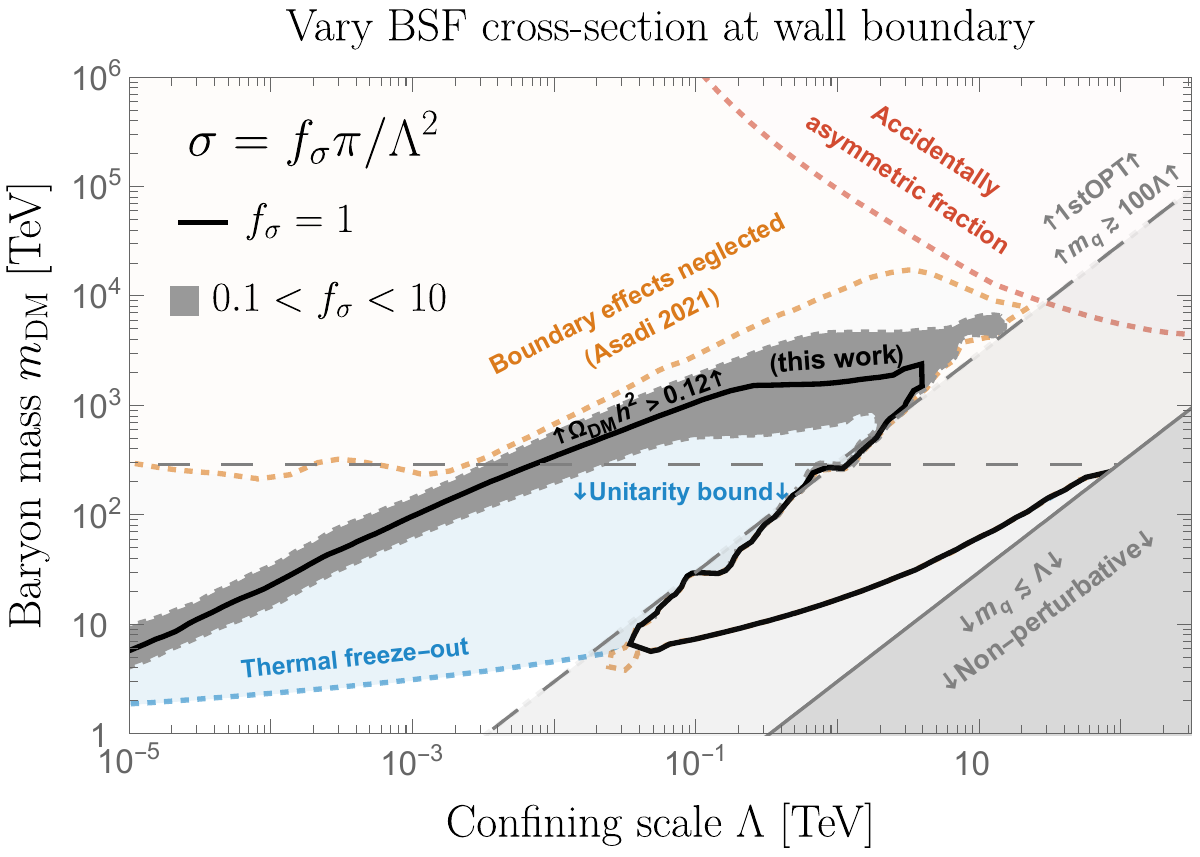}
    \caption{The \textbf{black line} shows where the dark baryon abundance, accounting for asymmetric and symmetric (bulk + boundary) contributions, satisfies the observed dark matter relic abundance. When decreasing the wall velocity below its maximal value $v_{\rm max}$ set by Hubble expansion in Eq.~\eqref{eq:v_w_contraction}, the symmetric component to the baryon abundance is depleted. In \textbf{bottom-right} panel, we vary the surface area $\sigma=f_\sigma \pi/\Lambda^2$, defined in Eq.~\eqref{eq:BSF_cross-section_wall}, of the slit on wall boundary through which dark quark can escape by forming a color-neutral bound-state. For all the panels, we remain agnostic about the portal to the SM; nevertheless, we assume a short glueball lifetime to prevent both entropy injection and the imposition of BBN constraints.}
    \label{fig:mDM_Lambda_vary}
\end{figure*}
\textbf{Bound-states formation before pocket formation.}
The bound-state formation rates have a strong dependence on the inverse radius, either at wall boundaries, $1/R^7$ or $1/R^4$ in previous equations, or in the bulk, $1/R^3$ in Eq.~\eqref{eq:finiteblz_app}. This implies that they only become relevant at late time after that the pocket has sufficiently contracted. Instead, during bubble expansion before pocket formation, for which $R\sim R_0$ in Eq.~\eqref{eq:R_0_def}, the formation of bound-states is very suppressed.\\
\textbf{Analytical Estimation.}
Only accounting for quark annihilation in the bulk and meson formation at the boundary, the Boltzmann equation for the quark density, $i=1$ in Eq.~\eqref{eq:boltz_app}, becomes
\begin{equation}
\label{eq:Rrec1}
v_w N_1'\simeq  \Gamma^{\rm bulk}_{(1,-1)\to(0,0)} + \Gamma_{(1,-1) \rightarrow \mathcal{M} }^{\rm bdy}  \simeq \frac{3\left<\sigma_{\rm ann} v\right>}{4\pi}\frac{N_1^2}{R^3}+ \frac{9f_\sigma v_{\mathcal{Q}}}{64 \Lambda^3}\frac{N_1^2}{R^4} ,
\end{equation} 
where we used $\dot{N}_1(t) = -v_wN_1'(R)$, Eqs.~(\ref{eq:finiteblz_app}, \ref{eq:1m1-meson}), $V=4\pi R^3/3$ and $\left<\sigma_{\rm ann} v\right> = \zeta  \pi \alpha_{\rm DC}^2/m_{\mathcal{Q}}^2$. As long as the quark annihilation rate in the bulk and at the boundary is smaller than the bubble expansion rate $v_w N_1/R$, the number of quark in the pocket is approximately constant. The number of quarks in the pocket starts to decrease substantially below a ``recoupling'' radius $R_{\rm rec}$:
 \begin{equation}\label{eq:RstarRb}
 	R_{\rm rec}=\textrm{Max}\left[R_{\rm bulk}, R_{\rm bdy} \right],
 \end{equation}
 where $R_{\rm bulk}$ and $R_{\rm bdy}$ are the radii when the first and second terms of the right hand side of Eq.~\eqref{eq:Rrec1} become of order $c_0 v_w N_1/R$ respectively, where $c_0\simeq 2.2$ is a parameter aposteriori fitted against numerical results:
 \begin{equation}\label{eq:RstarRb}
 	R_{\rm bulk}=\sqrt{3\frac{N^{\rm ini}_1\left<\sigma_{\rm ann} v\right>}{4\pi c_0v_w}}\;, \quad R_{\rm bdy}=\frac{1}{\Lambda}\left[\frac{9 f_\sigma v_{\mathcal{Q}} N^{\rm ini}_1}{64 c_0 v_w}\right]^{1/3}.
 \end{equation}
 Boundary terms are negligible in the regime $R_{\rm bdy} \ll R_{\rm bulk}$, which implies
 \begin{equation}
     v_w~ \lesssim~ \frac{64}{3\pi^3 c_0 f_\sigma^2 v_{\mathcal{Q}}^2} N_1^i\Lambda^6\left<\sigma_{\rm ann} v\right>^3.
 \end{equation}
Using Eqs.~\eqref{eq:Yqinf}, \eqref{eq:alpha_D_rge}, \eqref{eq:R_0_def} and \eqref{eq:quark_abundance_initial}, we obtain:
  \begin{equation}
  \label{eq:ve_bulk_dominate}
     v_w~ \lesssim~  \frac{2 \times 10^{-6}}{c_0f_\sigma^2}\left(\frac{\zeta}{0.26}\right)^2\left(\frac{\alpha_{\rm DC}}{0.08}\right)^4\left(\frac{g_{*SM}}{106.75}\right)^{1/2}\left(\frac{10^3\Lambda}{m_\mathcal{Q}}\right)^4\left(\frac{\rm TeV}{\Lambda}\right)^{1.7}.
 \end{equation}
Comparing with Eq.~\eqref{eq:pocket_velocity_app}, we deduce that boundary effects must be a priori included in the analysis. However, Eq.~\eqref{eq:pocket_velocity_app} is estimated as an upper limit of a more realistic pocket wall velocity accounting for inhomogeneity in the latent heat injection and possibly pressure contribution from heavy quarks \cite{Asadi:2021pwo}. For this reasons, in Fig.~\ref{fig:mDM_Lambda_vary}, we show the evolution of the DM parameter space when we vary the wall velocity $v_w$. We find indeed that boundary effects become negligible for $v_w \lesssim 10^{-4}v_{\rm max}$, with $v_{\rm max}$ given by Eq.~\eqref{eq:pocket_velocity_app}. In the same Fig.~\ref{fig:mDM_Lambda_vary}, we also show the impact of the uncertainties on the bound-state formation cross-section encoded in $f_\sigma$.
The integral in Eq.~\eqref{eq:S_symm_def_app} is dominated by the contribution around the recoupling radius. For $0<\lambda_2\ll \lambda_1\ll 1$, the survival factor can be estimated as:
\begin{equation}
\label{eq:surv_S_symm}
    S_{\rm sym}\simeq \left.\frac{R}{ v_w}\left(\Gamma_{3 \rightarrow 3}^{\rm esc} + 3\Gamma_{(1,1,1) \rightarrow 3}^{\rm bdy}\right)\right\vert_{R=R_{\rm rec}}\simeq \frac{3v_{\mathcal{Q}}}{4 v_w}\left(\frac{N_3}{N_1^i} + \frac{3 f_\sigma^2 }{512} \frac{(N_1^i)^2}{(R_{\rm rec}\Lambda)^6}\right) .
\end{equation}
 In Fig.~\ref{fig:ana_vs_num}, we successfully compare the analytical formulae in Eq.~\eqref{eq:surv_S_symm} to the numerical results after fixing $c_0\simeq 2.2$ in Eq.~\eqref{eq:RstarRb}. The baryon number $N_3$ in Eq.~\eqref{eq:surv_S_symm} is calculated from algebraically solving the simple system of Eqs.~\eqref{eq:quasi_eq_1} and \eqref{eq:quasi_eq_2}.\\
\textbf{Neglecting boundary effects.}
Before recoupling $R\gtrsim R_{\rm bulk}\gg R_{\rm bdy} $, the di-quark and baryon numbers $N_2$ and $N_3$ in Eqs.~\eqref{eq:boltz_app_N2} and \eqref{eq:boltz_app_N3}, follow the quasi-equilibrium conditions:
\begin{align}
    &\Gamma^{\rm bulk}_{(1,1)\to(2,0)}  \simeq\Gamma_{(2,1) \rightarrow (3,0)}^{\rm bulk}+ \Gamma_{(2,-1) \rightarrow (1,0)}^{\rm bulk} 
    ,\\
    &\Gamma^{\rm bulk}_{(2,1)\to(3,0)} \simeq \Gamma^{\rm bulk}_{(3,-1)\to(2,0)}
\end{align}
in which we have neglected boundary effects.
Using Eq.~\eqref{eq:finiteblz_app}, we rewrite those conditions as:
\begin{align}
\label{eq:quasi_eq_1}
\langle \sigma v\rangle_{11,20}\left(N_1^2-\tilde{f}_1 N_2 V\right)&=\langle \sigma v\rangle_{21,30}\left(N_2N_1-\tilde{f}_2 N_3 V\right)+\langle \sigma v\rangle_{2(-1),10}\left(N_2N_1-n_2^{\rm eq}  N_1 V\right)\;,\\
\label{eq:quasi_eq_2}
\langle \sigma v\rangle_{21,30}\left(N_2N_1-\tilde{f}_2 N_3 V\right)&=\langle \sigma v\rangle_{3(-1),20}\left(N_3N_1-\tilde{f}_3 N_2 V\right),
\end{align}
where:
\begin{align}
	&\tilde{f}_1=n_0f_{11,20}=N_{\rm DC}\left(\frac{m_{\mathcal{Q}}\Lambda}{\pi}\right)^{3/2}\exp\left(-\frac{E_2}{\Lambda}\right)\;,\quad \\
	&\tilde{f}_2=n_0f_{21,30}=2N_{\rm DC}^2\left(\frac{m_{\mathcal{Q}}\Lambda}{3\pi}\right)^{3/2}\exp\left(-\frac{E_3-E_2}{\Lambda}\right)\;,\quad \\
	&\tilde{f}_3=n_0f_{3-1,20}=\left(\frac{3m_{\mathcal{Q}}\Lambda}{\pi}\right)^{3/2}\exp\left(-\frac{2m_{\mathcal{Q}}+E_2-E_3}{\Lambda}\right),
\end{align}
where $f_{\rm abcd}$ are defined in Eq.~\eqref{eq:f_abcd}.
The binding energy $E_{2,3}$ reads \cite{Mitridate:2017oky}: 
\begin{equation}
	E_2=\frac{1}{16}C_N^2\alpha_{\rm DC}^2m_{\mathcal{Q}},\qquad
	\quad E_3=0.26 C_N^2 \alpha_{\rm DC}^2m_{\mathcal{Q}},
\end{equation}
where $C_N = (N_{\rm DC}^2-1)/2N_{\rm DC} = 4/3$ for $N_{\rm DC}=3$.
In the limit of efficient bound state breaking $\tilde{f}_k V \gg 1$, the diquarks and the baryons reach the chemical equilibrium:
\begin{equation}
	N_2\simeq \frac{N_1^2}{\tilde{f}_1V}\;,\quad N_3\simeq \frac{N_1^3}{\tilde{f}_1\tilde{f}_2V^2}.
\end{equation}
Plugging back into the survival abundance formula in Eq.~\eqref{eq:surv_S_symm} whose second term is neglected, we obtain:
\begin{equation}
\label{eq:surv_S_symm_bulk}
    S_{\rm bulk}\simeq \frac{3v_{\mathcal{Q}}}{4 v_w}\frac{(N_1^i)^2}{\tilde{f}_1\tilde{f}_2V_{\rm bulk}^2} \simeq c_0 \frac{\pi v_w^2}{\tilde{f}_1\tilde{f}_2N_1^i\left<\sigma_{\rm ann} v\right>^3}\sqrt{\frac{\Lambda}{m_\mathcal{Q}}},
\end{equation}
where $V_{\rm bulk}=4\pi R_{\rm bulk}^3/3$ with $R_{\rm bulk}$ the radius defined in Eq.~\eqref{eq:RstarRb} when quarks annihilation in the bulk recouple, and $N_1^i$ is the initial nunber of quarks per pocket in Eq.~\eqref{eq:quark_abundance_initial}.\\
\textbf{Neglecting bulk effects.}
We now assume bulk effects to be negligible and consider the phase before recoupling $R\gtrsim R_{\rm bdy} \gg R_{\rm bulk} $. Only accounting for the second term in Eq.~\eqref{eq:surv_S_symm} leads to the survival abundance: 
\begin{equation}
\label{eq:surv_S_symm_bdy}
    S_{\rm bdy} \simeq \frac{3v_{\mathcal{Q}}}{4c v_w} \frac{3 f_\sigma^2 }{512} \frac{(N_1^i)^2}{(R_{\rm bdy}\Lambda)^6} = \frac{2}{9}v_w\sqrt{\frac{m_{\mathcal{Q}}}{\Lambda}},
\end{equation}
where $R_{\rm bdy}$ the radius defined in Eq.~\eqref{eq:RstarRb} when quarks annihilation at the boundary recouples, and where we replaced $v_{\mathcal{Q}}\simeq \sqrt{\Lambda/m_{\mathcal{Q}}}$.
The baryon abundance after the PT reads 
\begin{equation}
\label{eq:Y_DM_after_PT}
    Y_{\rm DM} = S\times \frac{2Y_{\mathcal{Q}}^{\rm FO}}{N_{\rm DC}},
\end{equation}
with $S=S_{\rm bulk} +S_{\rm bdy} +S_{\rm asym} $, the factor $2$ accounting for anti-baryons. In the next section we discuss the possiblity that the dark baryon DM abundance is diluted by entropy injection following glueball decay.

\subsection{Baryon dilute-out}
\label{app:quark_DO}
Ref.~\cite{Asadi:2022vkc} shows that 3-to-2 interactions are efficient enough to maintain glueballs at thermal equilibrium with themselves right after the end of the PT, relaxing their chemical potential to zero, as long as $\Lambda \lesssim 10^5~\rm TeV$. Assuming that that glueballs are at thermal equilibrium with themselves, and at the same temperature as the SM (anyway glueballs can not be hotter than $ T_c$ \cite{Asadi:2022vkc} ), their abundance reads:
\begin{equation}
R_{\rm GB} \equiv \frac{s_{\rm GB}}{s_{\rm SM}} = \frac{45(2J+1)}{2\pi^2(2\pi)^{3/2}g_{*,\rm SM}} \left(\frac{m_{\rm GB}}{\Lambda} \right)^{5/2} e^{-m_{\rm GB}/\Lambda} \simeq 
        1.6\times 10^{-4},
\end{equation}
where $J$ is the glueball spin, taken here to $J=0$ for the lightest glueball state, and where we set $g_{*,\rm SM}=106.75$.
As the universe cools down, the glueball undergo a phase of cannibalism \cite{Carlson:1992fn,Pappadopulo:2016pkp} during which $3$-to-$2$ interactions reduce the number density which becomes more and more Boltzmann-suppressed until the rate of 3-to-2 interaction freezes-out below a temperature $x_{\rm FO} =m_{\rm GB}/T_{\rm FO}$ solution of \cite{Forestell:2016qhc}:
\begin{equation}
\label{eq:T_FO_GB_app}
x_{\rm FO}^{5/2} e^{2x_{\rm FO}} = \frac{g_*^{1/4}}{180\pi}R_{\rm GB}\left(m_{\rm GB}^4M_{\rm pl}\left<\sigma_{32}v^2\right>\right)^{3/2},
\end{equation}
where $\left<\sigma_{32}v^2\right>$ is the thermally averaged 3-to-2 cross-section \cite{Forestell:2016qhc}:
\begin{equation}
\left<\sigma_{32}v^2\right> \simeq \frac{1}{(4\pi)^3}\left( \frac{4\pi}{N_{\rm DC}} \right)^6 \frac{1}{m_{\rm GB}^5}.
\end{equation}
For $\Lambda \simeq \rm 100~\rm GeV$, we obtain $x_{\rm FO}\sim 18$ which implies that glueballs freeze-out rather shortly after the completion of the phase transition.
The resulting frozen-out relic abundance of glueballs $Y_{\rm GB} \equiv {n_{\rm GB}/s_{\rm SM}}$ is:
\begin{equation}
\label{eq:YGB_app}
Y_{\rm GB} \simeq \frac{R_{\rm GB}}{x_{\rm FO}}.
\end{equation}
If the glueball is long-lived, it can dominate the energy density of the universe below the temperature:
\begin{equation}
\label{eq:Tdom_app}
T_{\rm dom} = \frac{4}{3}m_{\rm GB}Y_{\rm GB}.
\end{equation}
The matter domination ends when glueballs decay, when the universe expansion rate crosses the decay rate $H\simeq \Gamma_{\rm GB}$, around the temperature:
\begin{equation}
\label{eq:Tdec_app}
T_{\rm dec}\simeq 1.3g_*^{1/4}\sqrt{\Gamma_{\rm GB} M_{\rm pl}}.
\end{equation}
During the decay, the non-relativistic glueballs degrees of freedom are converted into relativistic SM degrees of freedom. This leads to an increase of the SM entropy by the ratio, e.g. \cite{Cirelli:2018iax}:
\begin{equation}
D = \frac{S_f}{S_i} \simeq 1+\left( \frac{T_{\rm dec}}{T_{\rm dec}^{\rm before}} \right)^3 \simeq 1+ \frac{T_{\rm dom}}{T_{\rm dec}},
\end{equation}
where the last equality can be derived from evolving matter and radiation between the two epochs. An important consequence is the dilution of DM abundance:
\begin{equation}
Y_{\rm DM}~ \simeq~ \frac{S}{D}\, \frac{2Y_{\mathcal{Q}}^{\rm FO}}{N_{\rm DC}},
\end{equation}
where $S<1$ is the survival factor in Eq.~\eqref{eq:survival_fac_app} and the factor of $2$ includes the contribution from anti-baryons.

\section{Communication with SM}
\label{app:communicate_to_SM}

\begin{figure*}[h!]
    \centering
\begin{minipage}[c]{0.48\textwidth}

{\begin{tikzpicture}
\begin{feynman}
\vertex (a1);
\vertex[above=0.8cm of a1] (c1){\(H\)};
\vertex[below=0.8cm of a1] (d1){\(H\)};
\vertex[right=1cm of a1] (a6);;
\vertex[right=1cm of a6] (a7);
\vertex[above=0.8cm of a7] (d2){\(\phi\)};
\vertex[below=0.8cm of a7] (c3){\(\phi\)};

    \diagram*{
      (d1) -- [scalar] (a6),
      (c1) -- [scalar] (a6),
      (a6) -- [scalar] (d2),
      (a6) -- [scalar] (c3),

    };

\end{feynman}
\end{tikzpicture}
}

\end{minipage}
\begin{minipage}[c]{0.48\textwidth}
{\begin{tikzpicture}
\begin{feynman}
    \vertex (a1) {\( \phi \)};
    \vertex[right=1.cm of a1] (a2);
    \vertex[right=1cm of a2] (a3);
    \vertex[above=0.5cm of a3] (a4) ;
    \vertex[below=0.5cm of a3] (a5) ;
    \vertex[right=1cm of a4] (b4){\( \mathcal{G}\)};
    \vertex[right=1cm of a5] (b5){\( \mathcal{G}\)};

    \diagram*{
       {[edges=fermion],
      },

      (a2) -- [scalar] (a1),
      (a4) -- [fermion, edge label'=\(\ \mathcal{Q}\)] (a2),
      (a2) -- [fermion, edge label'=\(\ \mathcal{Q}\)] (a5),
      (a5) -- [fermion, edge label'=\(\ \mathcal{Q}\)] (a4),
      (a4) -- [boson] (b4),
      (a5) -- [boson] (b5),

    };

\end{feynman}
\end{tikzpicture}
}
\end{minipage}
    \caption{\textbf{Higgs portal}: A scalar $\phi$ has mixing with Higgs (\textbf{left} diagram) and Yukawa coupling to heavy quarks (\textbf{right} diagram). }
    \label{fig:heavY_quark_portal}
\end{figure*}
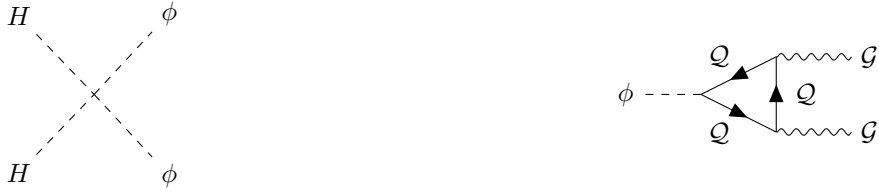
\subsection{Higgs portal}
\label{app:heavY_quark_portal}

We introduce a singlet scalar $\phi$ with mixing to SM Higgs $H$ and Yukawa coupling to heavy $SU(N_{\rm DC})$ quark $\mathcal{Q}$, see Fig.~\ref{fig:heavY_quark_portal}:
\begin{equation}
\label{eq:UV_lagrangian_Higgs_portal}
    \mathcal{L}\supset -\lambda_{H\phi}|H|^2|\phi|^2 + y_{\phi\mathcal{Q}}\phi \overline{\mathcal{Q}}\mathcal{Q}.
\end{equation}
We assume that the singlet scalar has a potential:
\begin{equation}
\label{eq:potential_phi}
    V(\phi) = \frac{\lambda_\phi}{4}\left(\phi^2-v_\phi^2\right)^2,
\end{equation}
giving it a vacuum expectation value (VEV) $v_\phi$, from which the heavy quarks get their mass $m_{\mathcal{Q}}=y_{\phi\mathcal{Q}} v_\phi/\sqrt{2}$.
Integrating the singlet $\phi$ and heavy quarks $\mathcal{Q}$ leads to a dimension 6 effective operator between SM Higgs and $SU(N_{\rm DC})$ gluons (see also \cite{Juknevich:2009gg}):
\begin{equation}
\label{eq:GB_op_dim_6}
    \mathcal{L}^{(6)} \supset  \frac{\alpha_{\rm DC}}{3\pi} \frac{\lambda_{H\mathcal{Q}}}{m_{\rm Q}^2} H^{\dagger}H \mathcal{G}_{\mu\nu}^A \mathcal{G}^{\mu\nu A},\qquad \textrm{with}\quad  \lambda_{H\mathcal{Q}} =\lambda_{H\phi} \frac{v_\phi m_\mathcal{Q}}{m_\phi^2}= \sqrt{2}\lambda_{H\phi} \left( \frac{m_{\mathcal{Q}}}{m_{\phi}} \right)^2,
\end{equation}
where the singlet scalar mass from Eq.~\eqref{eq:potential_phi} is $m_\phi^2 = 2\lambda_\phi v_\phi^2$.
 For the gauge group $SU(3)$, the lightest bound state in the dark sector is the glue-ball with quantum number $J^{PC}=0^{++}$ and mass $m_{0^{++}}\simeq 7 \Lambda$ \cite{Mathieu:2008me}. The annihilation matrix element of the $0^{++}$ is identified as the $0^{++}$ decay constant \cite{Juknevich:2009gg,Chen:2005mg}:
\begin{equation}
    F_{0^+}^S\equiv \langle 0|\mathcal{G}^a_{\mu\nu}\mathcal{G}^{a\mu\nu}|0^{++}\rangle.
\end{equation}
Lattice simulations of $SU(3)$ Yang-Mills theory give the relation~\cite{Chen:2005mg}:
\begin{equation}
    4\pi\alpha_{\rm DC} F_{0^+}^S\simeq 3 m_{0^{++}}^3,\qquad \textrm{with}\quad  m_{0^{++}}\simeq 7 \Lambda.
\end{equation}
In the rest, we denote $\mdg \equiv m_{0^{++}}$.
Applying Feynmann rules to Eq.~\eqref{eq:GB_op_dim_6}, one calculates the decay width of the lightest glueball state $0^{++}$ into Higgs pairs \cite{Juknevich:2009gg}:
\begin{equation}
\label{eq:GB_dec_dim_6_Higgs}
    \Gamma_{{\rm GB}\to hh} \simeq \frac{1}{32\pi m_{\rm GB}}\left(\frac{\lambda_{H\mathcal{Q}} \alpha_{\rm DC}F_{0^+}^S}{3\pi m_{\rm Q}^2} \right)^2,
\end{equation}
but also into $WW$ and $ZZ$ and SM fermions \cite{Juknevich:2009gg}:
\begin{equation}
\label{eq:GB_dec_dim_6_WW_ZZ}
    \Gamma_{\rm GB\to WW} =2 \Gamma_{\rm GB\to ZZ} = 2 \Gamma_{\rm GB\to hh},\qquad 
    \Gamma_{\rm GB\to f\overline{f}} = N_c \frac{4m_f^2}{m^2_{\rm GB}}   \Gamma_{{\rm GB}\to hh},
\end{equation}
where $N_c=3$ and $1$ for quarks and leptons. Note that we have omitted the mass threshold factors, implying that all decay widths $\rm GB \to XX$ must be supplemented by the conditions $m_{\rm GB}> 2m_X$.
We deduce the glueball lifetime for decay into Higgs and bottom quarks :
\begin{equation}
\tau_{\rm DG} \simeq 
\begin{cases}
10^{-6}~{\rm s}\left( \dfrac{0.01}{\lambda_{H\mathcal{Q}}} \right)^2\left( \dfrac{\rm TeV}{m_{\rm GB}} \right)^5\left( \dfrac{m_{\mathcal{Q}}}{10^3\rm TeV} \right)^4, \qquad\qquad \quad \qquad ~\qquad\qquad  \rm GB \to hh, \quad \textrm{if } m_{\rm GB}>2m_h\vspace{0.2cm}\\
5\times 10^{-5}~{\rm s}\left( \dfrac{3}{N_c} \right)\left( \dfrac{0.01}{\lambda_{H\mathcal{Q}}} \right)^2\left(\dfrac{m_b}{m_f} \right)^2\left( \dfrac{10~\rm GeV}{m_{\rm GB}} \right)^3\left( \dfrac{m_{\mathcal{Q}}}{10~\rm TeV} \right)^4, \qquad  \rm GB \to b\bar{b}, \quad \textrm{if } m_{\rm GB}>2m_b.
\end{cases}
\end{equation}
The CP even scalar glueball state $0^{++}$ with mass $m_{0^{++}}\simeq 7\Lambda$ is only the lightest among many other glueball states whose fate must be investigated. The lightest ones are $2^{++}$, $0^{-+}$, $1^{+-}$ and $2^{-+}$ with masses $m_{2^{++}}\simeq 1.39 m_{0^{++}}$, $m_{0^{-+}}\simeq 1.50 m_{0^{++}}$, $m_{1^{+-}}\simeq 1.70 m_{0^{++}}$ and $m_{2^{-+}}\simeq 1.79m_{0^{++}}$. Apart from $0^{-+}$ and $1^{+-}$, all the glueball state $J$ can decay radiatively $J\to J'h$ into a lighter glueball state $J'$ by emitting a Higgs boson $h$ with a decay width comparable to Eq.~\eqref{eq:GB_dec_dim_6_Higgs} \cite{Juknevich:2009gg}. Instead, $0^{-+}$ and $1^{+-}$ can decay via dimension 8 operators, $C$ and $P$ odd respectively
\begin{align}
\label{eq:EFT_C_odd}
       &\mathcal{L}^{(8)}_{\rm C~odd} \supset \left[c_8^{(1)}\,\textrm{Tr}\left(\mathcal{G}_{\mu\nu}\mathcal{G}_{\alpha\beta}\mathcal{G}^{\alpha\beta}\right)+c_8^{(2)}\,\textrm{Tr}\left(\mathcal{G}^\alpha_{\mu}\mathcal{G}^{\beta}_\alpha\mathcal{G}_{\beta\nu} \right)\right]F^{\mu\nu},
       \\
       \label{eq:EFT_P_odd}
       &\mathcal{L}^{(8)}_{\rm P~odd} \supset c_8^{(3)} \,\textrm{Tr}\left(\mathcal{G}_{\alpha\beta}\tilde{\mathcal{G}}^{\alpha\beta} \right)\textrm{Tr}\left(F^{\gamma \delta}\tilde{F}_{\gamma \delta}\right).
\end{align}
where $F_{\mu\nu}$ are field strengths of the SM gauge bosons. Since the field introduced in the UV Lagrangian in Eq.~\eqref{eq:UV_lagrangian_Higgs_portal} are all CP even, the operators in Eqs.~\eqref{eq:EFT_C_odd} and \eqref{eq:EFT_P_odd} can not be generated. Instead, they could be generated if the heavy quarks are directly charged under SM gauge groups \cite{Juknevich:2009ji,Juknevich:2009gg}. Keeping heavy quark neutral under SM gauge groups, the resonances $0^{-+}$ and $1^{+-}$ are stable.
The relic abundance of those two glueball states has been studied thoroughly in \cite{Forestell:2016qhc,Forestell:2017wov,Soni:2017nlm}, as a function of the ratio $R_{\rm GB}=s_{\rm GB}/s_{\rm SM}$ of the entropy stored in the dark sector to the entropy within after $SU(N_{\rm DC})$ confinement. Under the assumption that the glueballs are in thermal equilibrium with themselves, and maintain the same temperature as the SM, their abundance reads:
\begin{equation}
\label{eq:_R_GB_STATES}
    R_{\rm GB} \simeq \frac{45(2J+1)}{2\pi^2(2\pi)^{3/2}g_{*,\rm SM}} \left(\frac{m_{\rm GB}}{\Lambda} \right)^{5/2} e^{-m_{\rm GB}/\Lambda} \simeq 
    \begin{cases}
        1.6\times 10^{-4},\qquad \textrm{for}\quad {J^{\rm PC}=0^{++}},\\
        1.3\times 10^{-5},\qquad \textrm{for}\quad  {J^{\rm PC}=0^{-+}},\\
        1.3\times 10^{-5},\qquad \textrm{for}\quad  {J^{\rm PC}=1^{+-}}.\\
    \end{cases}
\end{equation}
In light of the value of $R_{\rm GB}$ in Eq.~\eqref{eq:_R_GB_STATES} and results from Ref.~\cite{Forestell:2016qhc}, we deduce that the relic abundance today of stable glueball states $0^{-+}$ and $1^{+-}$ is below the observed DM level if $\Lambda \lesssim 100~\rm TeV$ and $\Lambda \lesssim 10~\rm TeV$, respectively. We conclude that the relevant parameter space for dark baryon DM in Fig.~\ref{fig:mDM_Lambda_EW_portal} is not concerned by the relic abundance of heavier glueball states.

The condition of not worsening the  gauge hierarchy problem would require $\lambda_{H\phi}\lesssim m_h^4/v_\phi^4$ leading to $\lambda_{H\mathcal{Q}} \lesssim y_{\phi\mathcal{Q}}^4(m_h/2m_{\mathcal{Q}})^2/\sqrt{2}\lambda_\phi$. We suppose that the hierarchy problem is solved by other means and we set $\lambda_{H\mathcal{Q}}=0.01$ in Fig.~\ref{fig:mDM_Lambda_EW_portal}.

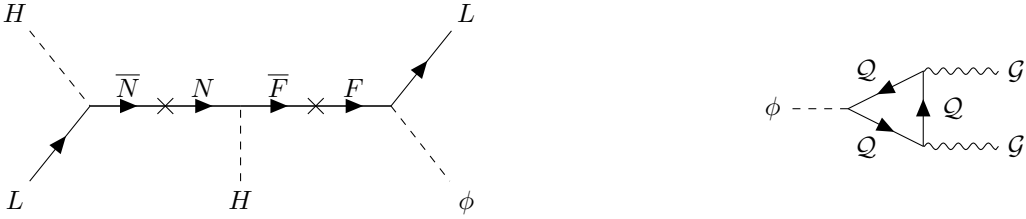
\begin{figure*}[h!]
    \centering
\begin{minipage}[c]{0.48\textwidth}
{\begin{tikzpicture}
\begin{feynman}
\vertex (a1);
\vertex[above=1cm of a1] (c1){\(H\)};
\vertex[below=1cm of a1] (d1){\(L\)};
\vertex[right=1cm of a1] (a2);
\vertex[right=1cm of a2] (a3);
\vertex[right=1cm of a3] (a4);
\vertex[right=1cm of a4] (a5);
\vertex[below=1cm of a4] (c2){\(H\)};
\vertex[right=1cm of a5] (a6);
\vertex[right=1cm of a6] (a7);
\vertex[above=1cm of a7] (d2){\(L\)};
\vertex[below=1cm of a7] (c3){\(\phi\)};

    \diagram*{
      (d1) -- [fermion] (a2),
      (c1) -- [scalar] (a2),
      (a2) -- [fermion,edge label=\(\overline{N}\)] (a3),
      (a3) -- [fermion,edge label=\(N\)] (a4),
      (a5) -- [fermion,edge label=\(F\)] (a6),
      (a4) -- [fermion,edge label=\(\overline{F}\)]  (a5),
      (a4) -- [insertion={[size=3pt]0.5}]  (a6),
      (a2) -- [insertion={[size=3pt]0.5}]  (a4),
      (c2) -- [scalar] (a4),
      (a6) -- [fermion] (d2),
      (a6) -- [scalar] (c3),

    };

\end{feynman}
\end{tikzpicture}
}
\end{minipage}
\begin{minipage}[c]{0.48\textwidth}
{\begin{tikzpicture}
\begin{feynman}
    \vertex (a1) {\( \phi \)};
    \vertex[right=1.cm of a1] (a2);
    \vertex[right=1cm of a2] (a3);
    \vertex[above=0.5cm of a3] (a4) ;
    \vertex[below=0.5cm of a3] (a5) ;
    \vertex[right=1cm of a4] (b4){\( \mathcal{G}\)};
    \vertex[right=1cm of a5] (b5){\( \mathcal{G}\)};

    \diagram*{
       {[edges=fermion],
      },

      (a2) -- [scalar] (a1),
      (a4) -- [fermion, edge label'=\(\ \mathcal{Q}\)] (a2),
      (a2) -- [fermion, edge label'=\(\ \mathcal{Q}\)] (a5),
      (a5) -- [fermion, edge label'=\(\ \mathcal{Q}\)] (a4),
      (a4) -- [boson] (b4),
      (a5) -- [boson] (b5),

    };

\end{feynman}
\end{tikzpicture}
}
\end{minipage}
    \caption{\textbf{Neutrino portal}: A light scalar $\phi$ couples to neutrino through the Dirac portal shown in \textbf{left} diagram and to glueballs through a loop of heavy quark as shown in  \textbf{right} diagram. }
    \label{fig:light_scalar_portal}
\end{figure*}

\subsection{Neutrino Portal}
\label{sec:light_scalar_portal}
We now assume that the singlet $\phi$ with Yukawa interaction with heavy quarks $\mathcal{Q}$ does not take a VEV anymore, and heavy quark masses are generated by other means. Instead of a mixing with Higgs it couples to the neutrino sector. More precisely, the single scalar $\phi$ couples to $SU(2)_L$ electronic doublet $L=(\nu_L,e_L)$, heavy vector-like $SU(2)$ doublet Dirac fermion $F,\bar{F}$, and neutral Dirac fermion $N, \bar{N}$, see Fig.~\ref{fig:light_scalar_portal}:
\begin{equation}\label{phiportal}
	\mathcal{L}=m_{\mathcal{Q}} \bar{\mathcal{Q}}\mathcal{Q}+\phi 
\bar{\mathcal{Q}}\left(y_{\phi\mathcal{Q}}+iy_{\phi\mathcal{Q},5}\gamma^5\right)\mathcal{Q}+\frac{m_\phi^2}{2}\phi^2+\lambda_1 LH\bar{N} +\lambda_2N H\bar{F}+\lambda_3\phi F L+M_1\bar{N} N+M_2\bar{F}F.
\end{equation}
 All the mass terms in Eq.~\eqref{phiportal} are of Dirac type and not Majorana. Consequently, the lepton number is conserved, and the SM neutrino remains massless \cite{Schmaltz:2017oov,Liu:2019ixm}. Integrating the heavy quarks at one loop leads to the effective operator:
\begin{equation}\label{eq:phiportal}
\mathcal{L}^{(5)}\supset \frac{\alpha_{\rm DC}\phi}{4\pi m_{\mathcal{Q}}}\left(y_{\phi\mathcal{Q}}\G\G+y_{\phi\mathcal{Q},5}\G\widetilde{\G}\right)+\frac12 y_{\phi\nu}\phi \,\overline{\nu}_L\nu_L,
\end{equation}
where $y_{\phi\nu}=\lambda_1\lambda_2\lambda_3\langle |H|^2\rangle/(M_1M_2)$. The decay of the lightest glueball state $0^{++}$ into neutrino via $s$-channel $\phi$-exchange $0^{++}\rightarrow \phi^*\rightarrow \nu\nu$ is given by the amplitude:
\begin{align}
\mathcal{M}&=\frac{\alpha_{\rm DC}y_{\phi\mathcal{Q}}y_{\phi\nu}}{4\pi m_{\mathcal{Q}}}\langle 0^{++}|\G^2\wick{\c1\phi|0\rangle\langle 0|\c1 \phi}\overline{\nu}_L\nu_L|\text{SM}\rangle\;.
\end{align}
We obtain glueball decay rate into neutrinos:
\begin{equation}
\label{eq:decay_width_GB_scalar}
\Gamma_{\rm GB}=\left(\frac{\alpha_{\rm DC}y_{\phi\mathcal{Q}}F_{0^+}^S}{4\pi m_{\mathcal{Q}}}\right)^2\frac{\Gamma_\phi}{\left(m_{\rm GB}^2-m_\phi^2\right)^2+m_\phi^2\Gamma_{\phi}^2}\;,\qquad 
{\rm GB}\to \nu\nu,
\end{equation}
where $\Gamma_\phi =   y_{\phi\nu}^2 m_\phi(1-4m_\nu^2/m_\phi^2)^{3/2}/8\pi$ is the $\phi$ decay rate into neutrinos. 
Existing constraints on the mediator $\phi$ coupling to neutrino arises from BBN, neutrinoless double beta decay, rare meson decay, neutrino scattering in DUNE, high-energy neutrinos in IceCube \cite{Berryman:2022hds}. In Fig.~\ref{fig:mDM_Lambda_scalar_portal}, we set $m_\phi \simeq 1~\rm GeV$, $y_{\phi\nu} \simeq 10^{-3}$ and $y_{\phi\mathcal{Q}}=1$ to circumvent all the aforementioned constraints while maximizing the decay width in Eq.~\eqref{eq:decay_width_GB_scalar} as a proof of principle that glueballs can have a shorter lifetime than via the Higgs portal explored in Sec.~\ref{app:heavY_quark_portal}.
We now discuss the fate of other light $J^{PC}$ states: $2^{++}$, $0^{-+}$, $1^{+-}$ and $2^{-+}$. The CP even state $2^{++}$ can radiatively decay to the lightest state by emitting a $\phi$ particle $2^{++}\to 0^{++}\phi$. The $P$ odd states $0^{-+}$ and $2^{-+}$ can radiatively transition to $P$ even states $J^{+-}\to J^{++}\phi$ thanks to the presence of the P violating term $y_{\phi\mathcal{Q},5}\bar{q}\gamma^5 q$ in Eq.~\eqref{phiportal}. As for the Higgs portal studied in Sec.~\ref{app:heavY_quark_portal}, the lightest $C$-odd state $1^{+-}$ is stable but its abundance is sub-dominant to the observed DM density if $\Lambda \lesssim 10~\rm TeV$ \cite{Forestell:2016qhc}.

\section{Phenomenology}
\label{app:pheno}

\subsection{Indirect detection}
Baryon can annihilate with anti-baryon in the galactic center where the dark matter density is at the highest. The cross-section reads \cite{Mitridate:2017oky,Geller:2018biy}:
\begin{equation}
\label{eq:sigma_rearr}
\left< \sigma_{\rm B\overline{B}} v_{\rm rel}\right> \sim \frac{\pi R_{\rm B}^2\left<v_{\rm rel}\right>}{\sqrt{E_{\rm kin}/E_{\rm B}}} \sim \frac{1}{\alpha_{\rm DC}m_{\mathcal{Q}}^2} \sim 10^{-24} \frac{\rm cm^3}{s}\left( \frac{10~\rm TeV}{m_{\rm DM}} \right)^2\left( \frac{0.1}{\alpha_{\rm DC}}\right),
\end{equation}
where $E_{\rm kin}=N_{\rm DC} m_{\mathcal{Q}} v_{\rm rel}^2/2$ is the kinetic energy, $R_B\simeq (\alpha_{\rm DC} m_{\mathcal{Q}})^{-1}$ is the baryon radius, $E_B \simeq 0.26 C_N^2 \alpha_{\rm DC}^2 m_{\mathcal{Q}}$ is the binding energy \cite{Mitridate:2017oky}, and $C_N = (N_{\rm DC}^2-1)/2N_{\rm DC} = 4/3$.
 At large velocity, $E_{\rm kin}\gg E_{\rm B}$, the quarks wave packet becomes smaller than the baryon size and the annihilation cross-section becomes the usual perturbative one $\left< \sigma_{\rm B\overline{B}} v_{\rm rel}\right> \sim \pi \alpha_{\rm DC}^2/m_{\mathcal{Q}}^2$. At low  velocities, $E_{\rm kin}\lesssim E_{\rm B}$, the annihilation cross-section is equal to the geometric cross-section fixed by the baryon radius $R_B$. The dominant annihilation channel is the re-arrangement of baryon and anti-baryon into unstable dark meson states that decay into SM particles.
 The branching ratio of mesons into SM particles in the final state is model dependent. We here consider two benchmark scenarios corresponding to the two SM portals introduced in App.~\ref{app:communicate_to_SM}:
\begin{enumerate}
    \item mesons decay solely to bottom quarks (Higgs portal),
    \item mesons decay solely to neutrinos (neutrino portal).
\end{enumerate} 
In the Higgs portal, we neglected that mesons can also decay into WW whose upper limits are slightly more constraining than with $bb$, and into $tt$ whose upper limits are almost indistinguishable from $bb$.
We use the limits from HESS~\cite{HESS:2016mib}, derived from $254$ hours of observation, assuming a standard NFW DM profile and local DM energy density $\rho_{\odot} = 0.39$~GeV/cm$^3$.
Ref.~\cite{Profumo:2017obk} already calculated the upper bound on the DM annihilation cross-section, assuming annihilation into an intermediate state, for DM mass up to $10~\rm TeV$. Following \cite{Cirelli:2018iax}, we use their limit on $\left<\sigma v\right>$, divided by a factor $2$ - because in the present case DM baryons are not self-conjugate - and extended up to $m_{\rm DM}=70$~TeV since the original HESS publication \cite{HESS:2016mib} provided limits up to that range.  The HESS constraints are shown in brown in Fig.~\ref{fig:mDM_Lambda_EW_portal}. For the neutrino portal, we assume that meson decay solely into electronic neutrino. We use the limits from ANTARES~\cite{Albert:2016emp}, derived assuming a NFW profile. The smearing of the neutrino energy spectrum by the one-step cascade is accounted by multiplying the upper bound on $\left<\sigma v\right>$ by a factor $\sim 2$ following \cite{Cirelli:2018iax}. We include an additional factor $2$ to account for dark anti-baryons. The ANTARES constraints are shown in brown in Fig.~\ref{fig:mDM_Lambda_scalar_portal}. See also \cite{Curtin:2022oec} for indirect detection signals assuming DM annihilation into dark glueballs.

\subsection{CMB}
DM annihilation during the era of recombination can inject additional energy into the SM bath, potentially altering the observed spectral characteristics of the CMB~\cite{Planck:2018vyg}. The efficiency of energy injection is quantified through efficiency factors, denoted as \(f^i_{\rm eff}(m_{\rm DM})\), and derived in~\cite{Slatyer:2015jla,Slatyer:2015kla}. Since these depend mostly on the total amount of energy injected, such limits do not depend on the number of steps between the DM annihilation and the final SM products~\cite{Elor:2015bho}.
Following \cite{Cirelli:2018iax}, we then place limits as follows:
\beq
\left< \sigma_{\rm B\overline{B}} v_{\rm rel}\right>   \, f^i_{\rm eff} < 8.2 \times 10^{-26} \,\frac{{\rm cm}^3}{\rm sec}\,\frac{m_{\rm DM}}{100~{\rm GeV}}\,,
\label{eq:CMB}
\eeq
where $\left< \sigma_{\rm B\overline{B}} v_{\rm rel}\right>$ given in Eq.~\eqref{eq:sigma_rearr} and $i=b\bar{b}$ for the Higgs portal scenario, see purple region in Fig.~\ref{fig:mDM_Lambda_EW_portal}. Instead, because of the evasive nature of neutrino, the neutrino portal scenario do not lead to any CMB constraints.

\subsection{Direct detection}
In the Higgs portal we consider, integration of the single scalar $\phi$ in Eq.~\eqref{eq:UV_lagrangian_Higgs_portal} leads to:
\begin{equation}
    \mathcal{L} \supset y_{\mathcal{Q}}\lambda_{H\phi}\frac{v_\phi}{m_\phi^2}  v_h h\overline{\mathcal{Q}}\mathcal{Q} -\sum_q \frac{m_q}{v_h}h\overline{q}q,
\end{equation}
where $q$ are SM quarks. In light of the coupling $\lambda_{H\mathcal{Q}}=\sqrt{2}\lambda_{H\phi}(m_{\rm Q}/m_\phi)^2$ introduced in Eq.~\eqref{eq:GB_op_dim_6}, the latter Lagrangian can be rewritten:
\begin{equation}
    \mathcal{L} \supset \lambda_{H\mathcal{Q}} \frac{v_h}{m_{\mathcal{Q}}}   h\overline{\mathcal{Q}}\mathcal{Q} -\sum_q \frac{m_q}{v_h}h\overline{q}q.
\end{equation}
Now integrating the Higgs boson, we get:
\begin{equation}
    \mathcal{L} \supset -\frac{\lambda_{H\mathcal{Q}}}{m_\mathcal{Q}}\frac{1}{m_{h}^2} \overline{\mathcal{Q}}\mathcal{Q} \sum_q m_q\overline{q}q.
\end{equation}
The resulting spin-averaged DM-nucleon cross-section is \cite{Ellis:2000ds,Agrawal:2010fh,Fedderke:2014wda}:
\begin{equation}
    \sigma_{\rm SI}^{N-\rm DM} = \frac{1}{\pi} \frac{\lambda_{H\mathcal{Q}}^2N_{\rm DC}^4}{m_{\rm DM}^2}\left( \frac{\mu_{N-\rm DM}f_N}{m_h^2}\right)^2,
\end{equation}
where $f_N =m_N\sum_{q=u,d,s}  \left(f_{T_q}^{(N)}+\frac{2}{9}f_{T_G}^{(N)}\right)\simeq 0.35 m_N$ is the nuclear
matrix element accounting for the quark and gluon content of the nucleon, $m_N\simeq 1~\rm GeV$ is the nucleon mass, and $\mu_{N-\rm DM}=m_{\rm DM}m_N/(m_{\rm DM}+m_N)$ is the reduced mass of the DM-nucleon system. The first piece of the color factor $N_{\rm DC}^4=N_{\rm DC}^2\times N_{\rm DC}^2$ comes from $m_{\rm DM}=N_{\rm DC}m_{\mathcal{Q}}$ at the denominator. The second color factor $N_{\rm DC}^2$ accounts for the coherent superposition of the $N_{\rm DC}$ heavy quarks in each DM particle (the typical exchanged $N-\rm DM$ momentum $\mu_{N-\rm DM}v \sim \rm MeV$ being much smaller than the inverse dark baryon radius $\alpha_{\rm DC}m_{\rm DM}$).  In the red region in Fig.~\ref{fig:mDM_Lambda_EW_portal}, we show XENON1T bounds \cite{XENON:2023cxc} assuming a rather large coupling $\lambda_{H\mathcal{Q}}=0.01$ (a choice which favour a weaker BBN constraints, see Sec.~\ref{app:BBN}). Instead, the neutrino portal is secluded from direct detection.

\subsection{Collider}
In principle, the Higgs portal would receive collider constraints on the Higgs mixing, e.g. \cite{Baldes:2022oev}
\begin{equation}
    \tan(2\theta_{H\phi})\simeq \frac{2\lambda_{H\phi}v_\phi v_H}{|m_\phi^2 -m_h^2|},
\end{equation}
at the level of $\theta_{H\phi} \lesssim 0.1$ for $m_\phi \sim \mathcal{O}(\rm TeV)$ and lighter \cite{Ilnicka:2018def}.
However the model possesses sufficient parameters, e.g. $v_\phi$, for $\theta_{H\phi}$ relevant for collider constraints and $\lambda_{H\mathcal{Q}}=\sqrt{\lambda_{H\phi}}m_{\mathcal{Q}}/m_\phi$ relevant for the glueball decay in Eq.~\eqref{eq:GB_op_dim_6}, to be independent quantities. For this reason, we do not investigate collider constraints further, and refer the reader to \cite{Mitridate:2017oky,Batz:2023zef} for dedicated studies.

\subsection{BBN}
\label{app:BBN}

\subsubsection{Glueball decay}

Glueballs are abundantly produced during the phase transition until they follow an equilibrium distribution. Glueballs then follow a short period of cannibalism until their abundance freezes-out, see App.~\ref{app:quark_DO}, and finally redshift like matter. Since they only communicate with the SM through high dimensional operators induced by loops of heavy quark, glueballs are long-lived. They can occupy a substantial fraction of the universe at the time of their decay, which is severely constraints by the successful prediction of BBN.

\textbf{Higgs portal.} In the Higgs portal, the glueball decays mostly hadronically.
The precise BBN constraints on such scenario have been thoroughly studied in \cite{Kawasaki:2017bqm}. We apply their results using the decay widths in Eqs.~(\ref{eq:GB_dec_dim_6_Higgs},\ref{eq:GB_dec_dim_6_WW_ZZ}) and the glueball abundance $Y_{\rm GB}$ in Eq.~\eqref{eq:YGB_app}. The resulting BBN constraint is shown in green in Fig.~\ref{fig:mDM_Lambda_EW_portal}.

\textbf{Neutrino portal.}
Particles decaying into neutrino are far less constrained than particles decaying into hadronic or electromagnetic products \cite{Ema:2014ufa,Hambye:2021moy}. As long as the glueball mass is smaller than $m_{\rm GB}\lesssim 0.1~\rm TeV$, it is a good approximation \cite{Hambye:2021moy} to simply account for the change in effective number of neutrino:
\begin{equation}
\label{eq:Delta_Neff_GB}
    \Delta N_{\rm eff} = \frac{8}{7}\left( \frac{\rho_{\rm GB}}{\rho_\gamma} \right)\left( \frac{11}{4} \right)^{4/3}\Big|_{T=T_{\rm dec}}= \frac{8}{7}\left( \frac{11}{4} \right)^{4/3} \frac{4}{3} \frac{m_{\rm GB}}{T_{\rm dec}}Y_{\rm GB},
\end{equation}
where $\rho_\gamma$ denotes the photon energy density, $T_{\rm dec}$ the photon temperature when glueballs decay, and $Y_{\rm GB}$ the glueball abundance in Eq.~\eqref{eq:YGB_app}.
CMB data \cite{ParticleDataGroup:2022pth} gives $N_{\rm eff} = 2.99_{-0.33}^{+0.34}$ and BBN predictions \cite{Mangano:2011ar, Peimbert:2016bdg} leads to $N_{\rm eff} = 2.90_{-0.22}^{+0.22}$. Instead, the predictions from the SM suggest \cite{Mangano:2005cc, deSalas:2016ztq} $N_{\rm eff}  \simeq 3.045$. The green region in Fig.~\ref{fig:mDM_Lambda_scalar_portal} shows the bound $\Delta N_{\rm eff}\lesssim 0.3$ in Eq.~\eqref{eq:Delta_Neff_GB}, using the decay width in Eq.~\eqref{eq:decay_width_GB_scalar}.
 
\subsubsection{Latent heat}
\label{app:latent_heat}
The latent heat $\LL$ of the PT contributes to the energy density of universe as a cosmological constant. More precisely, we can see from Eq.~\eqref{eq:vw_instant_therm} that in the regime of instantaneous reheating, the temperature of the universe is kept constant during the completion of the phase transition, in agreement with the interpretation of the latent heat as a vacuum energy.
The latent heat $\LL$ of the PT can modify the universe rate of expansion. This effect can be encapsulated within the effective number of neutrino relics as demonstrated below:
\begin{equation}
\label{eq:Neff_constraints}
\Delta N_{\rm eff} = \frac{8}{7}\left( \frac{\rho_{\rm tot}-\rho_{\nu}-\rho_{\gamma}}{\rho_\gamma} \right)\left( \frac{11}{4} \right)^{4/3} \simeq  6.7 \frac{\LL}{T_\gamma^4},
\end{equation}
where $\rho_{\gamma}$ and $\rho_{\nu}$ denotes the energy density of photons and neutrinos. 
The BBN restriction $ \Delta N_{\rm eff}  \lesssim 0.3 $ \cite{Pitrou:2018cgg,Dvorkin:2022jyg} is applicable post neutrino decoupling below the temperature $T_{\rm dec}\simeq 1~\rm MeV $ where $g_*(T< T_{\rm dec})\equiv 2+(7/8)\cdot 6\cdot (4/11)^{4/3} \simeq 3.36 $. 
Two distinct scenarios need to be considered separately.
\begin{enumerate}
    \item If the PT is reheated in the dark sector, then Eq.~\eqref{eq:Neff_constraints} and $\LL\simeq 1.413 T_{\rm D}^4$ for $N_{\rm DC}=3$ imply ($T_c\simeq \Lambda$):
    \begin{equation}
        \frac{T_{\rm D}}{T_{\gamma}}~\lesssim~0.42 \left(\frac{\Delta N_{\rm eff}}{0.3}\right).
    \end{equation}
    \item If the PT is reheated in the SM sector, $T_{\rm D}=T_{\gamma}$, then we get instead $N_{\rm eff} \simeq 9.5$, implying that we must enforce PT reheating temperature to be above the neutrino decoupling temperature, $ T_c \gtrsim 1~\rm MeV $. 
\end{enumerate}
The latter scenario is the focus of our study. It is worth noting that more stringent BBN restrictions have been discussed, such as $ T_{c} \gtrsim 3~\rm MeV $ \cite{Bai:2021ibt}, or even $ T_{c} \gtrsim 2~\rm MeV $ and $ 4~\rm MeV $ in the context of electromagnetic and hadronic decays respectively \cite{Kawasaki:2000en,Hasegawa:2019jsa}. In this work, we get rid  of regions associated with latent heat injection during BBN by cutting all our plots below $\Lambda \lesssim 10~\rm MeV$.

\bibliographystyle{apsrev4-1_bis}
\bibliography{squeezecharge}

\end{document}